\documentclass{aastex63}
\usepackage[utf8]{inputenc}
\usepackage[english]{babel}
\usepackage{graphics}
\usepackage{pgfplotstable}
\usepackage{natbib}
\usepackage{xspace}
\usepackage{booktabs}
\usepackage{longtable}
\usepackage{morefloats}
\usepackage{blindtext}
\usepackage{placeins}
\usepackage{makecell}
\usepackage[shortlabels]{enumitem}
\usepackage[caption=false]{subfig}
\usepackage{gensymb}
\setcitestyle{authoryear,open={(},close={)}}

\usepackage{float}
\submitjournal{AJ}

\shorttitle{The X-ray fundamental plane for classes}
\graphicspath{{./}{figures/}}
\begin{document}

\title{The X-ray fundamental plane of the Platinum Sample, the Kilonovae and the SNe Ib/c associated with GRBs}

\correspondingauthor{Dainotti, M. G.}
\email{mdainotti@stanford.edu}

\author{Dainotti, M. G.}

\affiliation{Interdisciplinary Theoretical \& Mathematical Science Program, RIKEN (iTHEMS), 2-1 Hirosawa, Wako, Saitama, Japan 351-0198}
\affiliation{Astronomical Observatory, Jagiellonian University, ul. Orla 171, 31-501 Krak{\'o}w, Poland; dainotti@oa.uj.edu.pl}
\affiliation{SLAC NATIONAL ACCELERATOR LABORATORY 2575 Sand Hill Road, Menlo Park, CA 94025, USA; dainotti@mailbox.slac.stanford.edu}
\affiliation{Department of Physics \& Astronomy, Stanford University, Via Pueblo Mall 382, Stanford, CA 94305-4060, USA; mdainott@stanford.edu}
\affiliation{Space Science Institute, Boulder, Colorado}

\author{Lenart, A.}
\affiliation{Faculty of Physics, Astronomy and Applied Computer Science, Jagiellonian University,  ul. prof. Stanisława Łojasiewicza 11, 30-348 Krak{\'o}w, Poland; aleksander.lenart@student.uj.edu.pl}

\author{Sarracino, G.}
\affiliation{Dipartimento di Fisica, ``E. Pancini''
Universit\`{a} ``Federico II'' di Napoli, 
Compl. Univ. Monte S. Angelo Ed. G, Via Cinthia, I-80126
Napoli (Italy)}
\affiliation{INFN Sez. di Napoli, Compl. Univ. 
Monte S. Angelo Ed. G, Via Cinthia, I-80126 Napoli (Italy)}

\author{Nagataki, S.}
\affiliation{RIKEN Cluster for Pioneering Research, Astrophysical Big Bang Laboratory (ABBL), 2-1 Hirosawa, Wako, Saitama, Japan 351-0198}
\affiliation{Interdisciplinary Theoretical \& Mathematical Science Program, RIKEN (iTHEMS), 2-1 Hirosawa, Wako, Saitama, Japan 351-0198}

\author{Capozziello, S.}
\affiliation{Dipartimento di Fisica, ``E. Pancini''
Universit\`{a} ``Federico II'' di Napoli, 
Compl. Univ. Monte S. Angelo Ed. G, Via Cinthia, I-80126
Napoli (Italy)}
\affiliation{INFN Sez. di Napoli, Compl. Univ. 
Monte S. Angelo Ed. G, Via Cinthia, I-80126 Napoli (Italy)}
\affiliation{Scuola Superiore Meridionale, Università di Napoli Federico II
Largo San Marcellino 10, 80138 Napoli (Italy)}

\author{Fraija, N.}
\affiliation{Instituto de Astronomia, Universidad Nacional Autonoma de Mexico, Apartado Postal 70264, C.P. 04510, Mexico D.F., Mexico}

\begin{abstract}
A large fraction of gamma-ray Bursts (GRBs) lightcurves (LCs) show X-ray plateaus. We analyze all GRBs with known redshifts presenting plateaus observed by The Neil Gehrels Swift Observatory from its launch until 2019 August. The fundamental plane relation between the rest-frame time and X-ray luminosity at the end of the plateau emission and the peak prompt luminosity holds for all the GRB classes when selection biases and cosmological evolutions are applied. 
We have discovered two important findings: (1) a new class of long GRBs with good data coverage: the platinum sample; and (2) the platinum, the SNe-LGRB and the KN-SGRB samples, yield  the  smallest  intrinsic  scatter with $\sigma_{platinum,GRB-SNe}=0.22 \pm 0.10$ and $\sigma_{KN-SGRB}=0.24 \pm 0.12$.
The SNe-LGRBs are composed of GRBs associated spectroscopically with the SNe Ib,c, the KN-SGRBs are composed by 8 GRBs associated with kilonovae or where  there could have been such an association.
The highest correlation coefficients are yielded for the SN-LGRB-ABC sample, which includes GRBs spectroscopically associated with SNe Ib/c or with a clear optical bump in the LC resembling the SNe Ib/c, ($R^{2}_{SN-LGRB-ABC}=0.95$), for the SN-LGRBs ($R^{2}_{SN-LGRB}=0.91$), and the KN-SGRBs ($R^{2}_{KN-SGRB}=0.90$) when the redshift evolution is considered. These category planes are reliable candidates to use as cosmological tools. Furthermore, the distance from the gold fundamental plane is a crucial discriminant among classes. In fact, we find that the distributions of the distances of the SNe-LGRB, SNe-LGRB-ABC, KN-SGRB and SGRB samples from the gold fundamental plane are statistically different from the distribution of the gold GRBs' distances from the gold fundamental plane with and without considering evolution cases. 
\end{abstract}

%In this analysis, we consider short GRBs, short GRBs with extended emission and intrinsically short (GRBs that have the intrinsic duration of the prompt emission $< 2$ s in the rest frame).


\section{Introduction} \label{sec:intro}
Gamma-ray Bursts (GRBs) are spectacular events, the most luminous panchromatic transient phenomena in the universe after the Big Bang, and are among the farthest astrophysical objects ever observed. One of the most challenging goals in modern astrophysics is the use of GRBs as standard candles. Their potential use as standard candles is similar to what has been done for SNe Ia, but GRBs are observed at much larger distances, allowing us to extend the cosmological ladder up to $z=9.4$. However, in order to use this approach, GRBs' emission mechanisms need to be very well understood. There is still an ongoing debate regarding their physical mechanisms and their progenitors. There are several proposed scenarios regarding their possible origin, e.g., the explosions of extremely massive stars and the merging of two compact objects, like neutron stars (NSs) and black holes (BHs). Both these models can have as central engines ordinary NSs, BHs, or fast spinning newly born highly magnetized NSs (magnetars). In the former scenario the compact object acting as the central engine is the remnant of the massive star after its collapse, while in the latter it is the result of the merging process and its subsequent explosion.

To pinpoint the different origins, we need to categorize GRBs according to their phenomenology. 
The GRB prompt emission is usually observed from hard X-rays to $\ge$ 100 MeV $\gamma$-rays, and sometimes also in optical wavelengths. The afterglow is the long-lasting multiwavelength emission (in X-rays, optical, and sometimes radio) following the prompt emission. 

\noindent GRBs are traditionally classified as short (SGRBs) and long GRBs (LGRBs), depending on the prompt emission duration: $T_{90}\leq 2$ s or $T_{90} \ge 2$ s,\footnote{$T_{90}$ is the time over which a burst emits from $5\%$ to $95\%$ of its total measured counts in the prompt emission.} respectively (Mazets et al. 1981, Kouveliotou et al. 1993).
A different classification based on physical mechanisms related to the GRBs' progenitors has been proposed (Zhang et al. 2009, Kann et al. 2011, Berger 2014, Li et al. 2020, Fraija et al. 2020), according to which GRBs are divided into Type I GRBs, which are powered by compact object mergers, and Type II GRBs, which have massive stars as progenitors. 

According to this classification Type I GRBs have the following features:
\begin{itemize}
    \item $T_{90}\leq 2 $ s.
    \item No SN association.
    \item They reside in elliptical or early-type galaxies, where generally no massive stars are found, and with low star formation rates (SFRs).
    \item They received a “natal kick” so that they are pushed away from their original birth site.
\end{itemize}
Candidates of Type I GRBs have a low density medium and small values for $E_{\gamma}$ and $E_{k}$,  which are the prompt emission isotropic energy corrected for the jet opening angle and the kinetic energy, respectively.

Type II GRBs have the following properties:
\begin{itemize}
    \item $T_{90}\geq 2 $ s and  $T_{90}/(1+z) \geq 2 $ s (the so-called intrinsic LGRBs).
    \item Clear SNe association.
    \item They reside in galaxies with high SFR.
    \item They explode in the same location where the progenitor stars are formed.
    \item A stratified stellar-wind-type medium ($n \propto R^2$, where $n$ is the density and $R^{2}$ is the radius of the  progenitor star) (Dai \& Lu 1998; Chevalier \& Li 2000).  
\end{itemize}

Candidates of Type II GRBs have high values for $E_{\gamma}$ and $E_{k}$.
%This classification resembles the one used for SNe and it does not depend only on the duration of the burst, but on multiple properties, like the wind properties produced by the afterglow and high values of the energy produced 
A diagram that clarifies this classification is shown in Figure 8 of Zhang et al. (2009).
The two classifications described above can be summarized by two main facts:
Type II GRBs are characterized by the collapse of massive stars (Woosley et al. 1993, the Collapsar), which means they should include the LGRBs, while Type I GRBs are characterized by %originate 
the merger of two NSs or an NS and a BH (Lattimer \& Schramm 1976, Narayan et al. 1992) and so SGRBs should belong to this class. However, also in this classification as well as in the morphological categorization, the correspondence between long/short and Type II/Type I GRBs is not universal: for instance, some SGRBs have been found belonging to the Type II class (Zhang et al. 2009).

In order to homogenize the morphological classification with the one that may arise from different progenitors or the same progenitors with different environments, we ascribe the morphological subclasses to the Type I or Type II categories.
%Such categories are
The categories that are comprised by Type II GRBs are LGRBs, the X-ray flashes (XRFs) with unusually soft spectra and greater fluence in the X-ray band (2-30 keV) than in the $\gamma$-ray band (30-400 keV), and ultra-long GRBs (ULGRBs) with a very long prompt duration ($T_{90}>1000$ s, Gendre et al. 2013, 2019; Piro et al. 2014). We consider here the cases that belong to the control sample in Gendre et al. (2019), where the end time of the prompt emission is measured until the beginning of the steep decay phase after the prompt), and the GRBs associated with supernovae (SN-LGRBs; Bloom et al. 1999). The categories associated with Type I GRBs are SGRBs, Short GRBs with extended emission (SEEs; Norris \& Bonnell 2006, Levan et al. 2007 and Norris et al. 2010) with mixed features between short and long GRBs, and GRBs associated with KNe (KN-SGRBs). Regarding the SN-LGRB and KN-SGRB categories, it could be possible that for most of them this association may not have been detected because of observational selection effects such as the Malmquist bias effect (Eddington 1913, 1940; Malmquist 1925) or due to the limited sensitivities of the observing satellites. In this regard, to better understand the role of selection biases on all these classes, we have treated them with the Efron \& Petrosian (EP, 1992) method. Although it is possible that the SN-LGRBs may not be physically distinct classes from LGRBs, it is still important to consider them as a different empirical subclass. Indeed, this segregation is essential because there are LGRBs for which an associated SN has not been detected even if it should have been clearly observed, e.g. the nearby $z = 0.09$ SN-less GRB 060505, and GRB 060614A, with $z = 0.125$, which could mean that further studies on the SN-less cases and their physical mechanisms are needed.  
Another class of GRBs is the intrinsically short (IS) GRB class with the rest frame $T^*_{90}=T_{90}/(1+z)<2$ s (Levesque et al. 2010), that we consider as a unique class with the SGRBs and SEEs. 
As pointed out in Dainotti et al. (2008, 2010, 2015a, 2016, 2017a, 2017b) and in Del Vecchio et al. (2016) for obtaining a class of GRBs that can be well standardized we need to select a subsample of GRBs with very well-defined properties from a morphological or a physical point of view.
We focus our attention mainly on KN-SGRBs and SNe-LGRBs as well as on the discovery of a platinum sample. This last sample is studied to fine tune more the classification of gold GRBs in an attempt to obtain the tightest possible plane and thus can be used as a cosmological tool.
The Neil Gehrels Swift Observatory (hereafter Swift) allows the observations of the X-ray plateau emission (O'Brien et al. 2006, Sakamoto et al. 2007 and Evans et al. 2009), which generally lasts from $10^2$ to $10^5$ s and is followed by a power-law (PL) decay phase.

Several models have been proposed to explain the plateau emission: the long-lasting energy injection into the external shock, where a single relativistic blast wave interacts with the surrounding medium (Zhang et al. (2006) and the spin-down luminosity of a magnetar (Stratta et al. 2018). The plateau emission is called external in the former case and internal in the latter one. The difference between these two origins can be derived from the value of the temporal power-law (PL) decay index of the plateau, $\alpha_{i}$: a very steep decay, $\alpha_{i} \ge 3$ for Li et al. (2018) and $\alpha_{i} \ge 4$ for Lyons et al. (2010), indicates the possible internal origin of the plateau (Willingale et al. 2007).

In \S \ref{sample selection} we describe the data samples, in \S \ref{Kilonovae} we summarize KNe observations, and in \S \ref{3D correlation} we present the three-parameter Dainotti relations and the results of the GRB samples, including the distributions of the distances of all the classes from the gold fundamental plane. In \S \ref{3D correlation with evolution} we present the fundamental plane relations for all the samples, correcting for selection biases and redshift evolution. We summarize and discuss our conclusions in \S \ref{discussion}.

\section{The sample Selection}\label{sample selection}
We analyzed comprehensively all GRBs presenting X-ray plateau afterglows detected by  Swift from 2005 January up to 2019 August with known redshifts,
spectroscopic or photometric, available in Xiao $\&$ Schaefer (2009), on the Greiner web page,\footnote{http://www.mpe.mpg.de/jcg/grbgen.html} and in the Gamma-ray Coordinates Network (GCN) circulars and
notices,\footnote{http://gcn.gsfc.nasa.gov/} excluding redshifts for which there is only a lower or an upper limit. More specifically, we have analyzed all 372 GRBs observed by Swift with known redshift from  2005 January observed up to 2019 August. The redshift range of our sample is $(0.033, 9.4)$. As shown in Dainotti et al. (2010), requiring an observationally homogeneous sample in terms of $T^{*}_{90}$ and spectral properties implies separating the sample into all the classes mentioned in the introduction. We gather Short, SEE, and   IS GRBs in one class, called hereafter the SGRB class. The ULGRBs of our sample have been chosen from the Gendre et al. (2019) samples (gold, silver and control, where there are 21 GRBs in total). We note here that one more ULGRB (091024) has been observed by Swift together with the Konus Wind (Virgili et al. 2013), that does not have a plateau, this is the reason why it does not belong in our sample. After the segregation in categories, our sample of 222 GRBs has been divided into:138 LGRBs, 20 XRFs, 22 SN-LGRBs, 43 SGRBs (12 IS, 14 SSE, and 17 Short), 11 ULGRBs, and 8 GRBs associated with KNe. We point out that the LGRB sample has been built from the whole sample, subtracting the SGRBs, XRFs, SN-LGRBs, ULGRBs, and KN-SGRBs, which means that a GRB belonging to the long class cannot be a part of the other classes mentioned here. Each GRB may belong to more than one empirical class, i.e., because all the GRBs associated with KNe are short, they will belong to both the KN-SGRB and SGRB categories.
We further classify the SN-LGRBs following Hjorth $\&$ Bloom (2011). The categories created are (A) strong spectroscopic evidence for an SN associated with the GRB; (B) a clear LC bump and some spectroscopic evidence suggesting the LGRB-SNe association; (C) a clear bump on the LC consistent with the LGRB-SNe associations, but no spectroscopic evidence of the SN; (D) a significant bump on the LC, but the properties of the SN are not completely consistent with other LGRB-SNe associations or the bump is not well sampled, or there is no GRB spectroscopic redshift; and (E) a bump, with low significance or inconsistencies with other observed LGRB-SNe identifications, but with the GRB spectroscopic redshift. The first three categories of this classification indicate a clear association of an SN event to an observed GRB, which allows us to create another subsample, called the SN-LGRB-ABC, formed by 14 GRBs. The whole SN-LGRB subsample is shown in Table \ref{TableSN}. Lastly, we have considered in our analysis the whole Type II GRB sample, which is composed of 179 GRBs, including internal plateaus. %26,186
\begin{table}
\centering
\hskip-3.0cm 
\caption{Table with $L_{peak}$, $T_{X}^{*}$, $L_{X}$ with Their Respective Errors, $z$ and the Classification According to Hjorth $\&$ Bloom (2011) of the 22 SN-LGRBs present in Our Sample.}
\begin{tabular}{ |c|c|c|c|c|c|c|c|c|c||c| } 
 \hline
 GRB & subclass & $L_{peak}$ $(erg/s)$ & $T_{X}^{*}$  $(s)$ & $L_{X} $ $(erg/s)$ & $z$ \\ \hline
 161219B & B & $49.31\pm 0.03 $ & $3.95 \pm 0.03$  & $45.62 \pm 0.03$ & 0.147  \\ 
 060707 & C & $51.60 \pm 0.15$  & $2.94 \pm 0.13$ & $48.07 \pm 0.10$ &  3.08 \\ 
 081007 & B & $50.29 \pm 0.05$ & $3.40 \pm 0.08$ & $46.44 \pm 0.06 $  & 0.529 \\
 090618 & C & $51.44 \pm 0.01$ & $3.485 \pm 0.014$ & $47.40 \pm 0.02 $  & 0.54  \\ 
 091127  & A & $51.41 \pm  0.02 $ & $3.81 \pm 0.02$ & $47.07 \pm 0.02$  & 0.49  \\ 
 060904B & C & $50.47 \pm 0.03$& $3.64 \pm 0.08$& $46.36 \pm 0.12 $ & 0.703 \\ 
 080319B & C & $51.73 \pm 0.03$ & $5.08 \pm 0.09$ &  $45.4  \pm 0.1$ & 0.937  \\ 
 101219B & B & $50.09 \pm 0.11$ & $4.23 \pm 0.17$ & $45.1 \pm 0.1$  &  0.552 \\
 120422A & A & $49.01 \pm 0.08$ & $5.13 \pm 0.22$ & $43.66 \pm 0.12$ & 0.28  \\
 130831A & B & $50.862 \pm 0.014 $ & $3.15 \pm 0.09$ & $47.05 \pm 0.06$ & 0.479  \\
 141004A & B & $50.691 \pm 0.014$ & $3.11 \pm 0.08$ & $46.52 \pm 0.09$ & 0.57   \\
 171205A & A & $47.26  \pm 0.09$ & $5.47 \pm  0.11$ & $42.06 \pm 0.07$ & 0.037  \\
 180728A & B & $50.474 \pm 0.004$ & $3.821 \pm 0.015$ & $46.12 \pm  0.01$ & 0.117  \\
 060218 & A & $46.08 \pm 0.09$ & $5.06 \pm 0.14 $ & $42.62 \pm 0.16$ & 0.033  \\
 090424 & E & $51.707 \pm 0.025$ & $2.81 \pm 0.01$ & $48.00 \pm 0.01$ & 0.544  \\
 100621A & E & $50.961 \pm 0.015$ & $3.45 \pm 0.06$ & $47.1 \pm 0.1$ & 0.542\\
 120729A & D & $50.69 \pm 0.04 $ & $3.27 \pm 0.05$ & $47.1 \pm 0.1$ & 0.8  \\
 050824 & E & $49.97 \pm 0.14$ & $4.82 \pm 0.13$ & $45.24 \pm  0.07$ & 0.83  \\
 051109B & E & $47.76 \pm 0.08$ & $3.62 \pm 0.13 $ & $43.6 \pm 0.1$ & 0.08  \\
 100418A & D & $50.1 \pm 0.1$ & $5.33 \pm 0.07$ & $44.7 \pm 0.1$ & 0.08  \\
 150821A & E & $53.15 \pm 0.14$ & $2.71 \pm 0.02$ & $48.53 \pm 0.02$& 0.755 \\
 060729 & E & $49.91 \pm 0.04$ & $4.918 \pm 0.013$ & $45.97 \pm 0.04$ & 0.54  \\
 \hline
\end{tabular}

 \textbf{Note.} All the values presented here but the redshift are in logarithm.
\label{TableSN}

\end{table}

We download the BAT + XRT LCs from the Swift web page repository. \footnote{http://www.swift.ac.uk/burst\texttt{\_}analyser}
We include all GRBs that can be fitted by the phenomenological Willingale et al. (2007, hereafter W07), model:

\begin{equation}
f(t) = \left \{
\begin{array}{ll}
\displaystyle{F_i \exp{\left ( \alpha_i \left( 1 - \frac{t}{T_i} \right) \right )} \exp{\left (
- \frac{t_i}{t} \right )}} & {\rm for} \ \ t < T_i \\
~ & ~ \\
\displaystyle{F_i \left ( \frac{t}{T_i} \right )^{-\alpha_i}
\exp{\left ( - \frac{t_i}{t} \right )}} & {\rm for} \ \ t \ge T_i, \\
\end{array}
\right .
\label{eq: fc}
\end{equation}

\noindent where both the prompt (index `i=\textit{p}') $\gamma$-ray and the initial X-ray decay and the afterglow (`i=\textit{a}') are modeled. The LC $f_{tot}(t) = f_{p}(t) + f_{a}(t)$ contains two sets of four free parameters $(T_{i},F_{i},
\alpha_{i}$, and $t_{i})$, where %$\alpha_{i}$ is the temporal PL decay index, and 
$t_{i}$ is the
initial rise timescale. We exclude cases %when 
when the afterglow fitting procedure fails or when the determination of 1 $\sigma$ confidence
intervals does not satisfy the Avni (1976) $\chi^{2}$ rules; see the XSPEC manual.\footnote{http://heasarc.nasa.gov/xanadu/xspec/manual/XspecSpectralFitting.html}  
We compute the source rest-frame isotropic luminosity $L_X$ and $L_{peak}$ (erg $s^{-1}$) in the Swift-XRT and BAT bandpass, $(E_{min}, E_{max})=(0.3,10)$ and $(E_{min}, E_{max})=(15,150)$ keV, respectively, as follows:
\begin{equation}
L_X= 4 \pi D_L^2(z) \, F_X (E_{min},E_{max},T^{*}_{X}) \cdot \textit{K}, \qquad L_{peak}= 4 \pi D_L^2(z) \, F_{peak} (E_{min},E_{max},T^{*}_{X}) \cdot \textit{K},
\label{eq: la}
\end{equation}
%et aket al.
where $D_L(z)$ is the luminosity distance, assuming a flat $\Lambda$CDM cosmological model with $\Omega_{M}=0.3$ and $H_{0}=70$ km s$^{-1}$ Mpc$^{-1}$ (Scolnic et al. 2018), $F_X$ and $F_{peak}$ are the measured $\gamma$-ray energy flux $($erg cm$^{-2}$ s$^{-1})$ at time $T_X$, the end of the plateau emission, and in the peak of the prompt emission over a $1$ s interval, respectively. \textit{K} is the \textit{K}-correction for cosmic expansion (Bloom et al. 2001). For  Swift-XRT GRBs, $\textit{K}=(1+z)^{(\beta-1)}$, where $\beta$ is the X-ray spectral index of the plateau phase. We derive the spectral parameters following Evans et al. (2009). For the prompt emission spectral fitting, we follow Sakamoto et al. (2011): when the $\chi_{CPL}^{2}-\chi_{PL}^{2}<6$, a PL or a cutoff power law (CPL) can be chosen, since the goodness of the fit is equivalent. We choose the CPL. We discard six GRBs that were better fitted with a blackbody model than with a PL and CPL. 
These requirements reduce the sample to $222$ GRBs. Finally, we construct a subsample of the LGRBs by including these morphology conditions: the beginning of the plateau should have at least five data points and the plateau should not be too steep (the angle of the plateau must be less than $41 \degree $). \footnote{The angle of the plateau is obtained as $\Delta_F/\delta_T=F_i-F_a/T_{X}-T_i$ using trigonometry, where $i$ is the time of the beginning of the plateau emission} This data quality criterion defines the gold sample, which includes $69$ GRBs (Dainotti et al. 2016, 2017a).

\subsection{Selection criteria for the Platinum Sample}
To further reduce the intrinsic scatter of the fundamental plane and increase its reliability as a cosmological probe, we define a subsample of the gold sample, the platinum sample. This is obtained after removing gold GRBs that present at least one of the following features:
\begin{itemize}
\item $T_{X}$ is inside a large gap of the data, and thus has a large uncertainty.
\item A small plateau duration $( <500$ s$)$ with gaps after it. This could mean that the plateau phase is longer than the one observed.
\item Flares and bumps at the start and during the plateau phase.
\end{itemize}

The LCs with these features create a sample of 50 platinum GRBs.
Lastly, we have segregated in  the internal plateau class the GRBs with internal plateaus according to Lyons et al. (2010) and Li et al. (2018) that belong to our sample (12 GRBs), for which the W07 model has problems in the fitting of the prompt phase. After this selection, the final sample is composed of a total number of 222 GRBs divided in the following way: 65 gold GRBs, 47 platinum GRBs, 129 LGRBs, 43 SGRBs, 22 SN-LGRB, 14 SN-LGRB-ABC, 18 XRFs, 10 ULGRBs, 8 KN-SGRBs, 167 Type II GRBs and 12 GRBs with internal plateaus as detailed in Table \ref{Table1}. 
More specifically, some of the XRFs are also in common with ULGRBs and SNe-LGRBs, thus the number of 167 does not include duplicates. There are 11 XRFs that are also SNe-LGRBs (two from the A and B classes, three from the C and E classes, and one from the D class) and one that is also an SNe-LGRB and an ULGRB.
We note here that 9 out of 12 GRBs of our sample have internal plateaus that belong to the LGRB, 1 belongs to the ULGRB and the remaining two to the XRF classes. Out of these 12, 10 have very high redshifts, with $z \geqslant 2.352$. We note, then, that Type I GRBs in our sample correspond to the SGRBs and that after we perform the segregation in Type I, II and internal plateaus, there is no overlap between these samples.

\section{The kilonovae in our sample compared to the AT 2017gfo Kilonova}\label{Kilonovae}

\begin{table}
 \centering
\hskip -3.0cm
\caption{Table with $L_{peak}$, $T_{X}^{*}$, $L_{X}$ with Their Respective Errors and $z$ of the Eight KN-SGRBs Present in Our Sample.}
\begin{tabular}{ |c|c|c|c|c|c|c|c|c|c|c| } 

 \hline
 GRB & $L_{peak}$ $(erg s^{-1})$ & $T_{X}^{*}$  $(s)$ & $L_{X} $ $(erg s^{-1})$ & $z$  \\ \hline
 060614A & $49.51 \pm 0.02$ & $4.98 \pm 0.03$ & $43.81 \pm 0.04$ & 0.125  \\ 
 061201 & $49.00 \pm 0.02$ & $3.45 \pm  0.09$ & $45.04 \pm 0.09$ &  0.111  \\ 
 070809 & $49.06 \pm 0.04$ & $4.1 \pm 0.2$ & $44.1 \pm 0.2$  & 0.219  \\
 070714B & $50.74 \pm 0.02$& $2.95 \pm 0.09$ & $46.9 \pm 0.1$  & 0.92 \\ 
 100625A  & $50.09 \pm 0.02$ & $2.28 \pm 0.24$ & $46.0 \pm 0.4$  & 0.452 \\ 
 111117A & $51.0 \pm  0.2$& $2.5 \pm  0.1$& $46.9 \pm 0.2$ & 2.21  \\ 
 130603B & $50.28 \pm 0.05$ & $3.40 \pm 0.05$ &  $46.1 \pm 0.2$  & 0.356 \\ 
 140903A & $49.80 \pm 0.03$ & $4.16 \pm 0.07$ & $45.25 \pm 0.06$ &  0.351 \\ 
 \hline
\end{tabular}

        \textbf{Note.} All values presented here except the redshift are in logarithm.
\label{TableKilo}

\end{table}

A careful analysis of the KNe properties %are
is
important given the discovery of the SGRB 170817A (Goldstein et al. 2017) associated with the AT 2017gfo KN (Coulter et al. 2017) and the detection of gravitational waves associated with this event. This observation
sheds light on the theoretical interpretation of SGRBs as compact NS mergers (Abbott et al. 2017). 

SGRBs are usually discovered through the detection of the $\gamma$-ray jet, which means that they are typically observed where the afterglow is brightest and so the KNe associated with them are more likely to be observed when the viewing opening angle is larger than the jet opening one (Metzger \& Berger 2012).
Here, we choose from our sample the GRBs associated with KNe present in the literature.
We aim to use the fundamental plane relation as a discriminant among the cases in which it is hard to verify if a KN could have been observed. If these uncertain cases of GRBs associated with KNe follow the KN-SGRB fundamental plane, then we can assert that these are associated with KNe.
In our sample of KN-SGRBs presenting plateaus we have the following GRBs: 060614A, 070714B, 130603B, 070809, 111117A, 140903A, 100625A, and 061201. Some of their physical parameters are presented in Table \ref{TableKilo}.

Gao et al. (2015, 2017) found four possible candidates for KN-SGRBs powered by a magnetar born after a merging event between two compact objects among 96 SGRBs observed by Swift that obey the following criteria: they have internal plateaus or extended emission, high-quality late-time data in both X-ray and optical bands, and redshift measurements. These are GRBs 080503, 050724, 070714B, and 061006. Among them, the SEE GRB 070714B belongs to our sample as well.

Gompertz et al. (2018) analyzed 23 nearby SGRBs ($z \leqslant 0.5$) to compare the optical and near-infrared LCs of the KN AT 2017gfo, to their counterparts to characterize the KNe diversity in terms of their brightness distribution.
The bursts that exclude the evidence of a KN similar to AT 2017gfo by several magnitudes can be a clue that a significant diversity exists in the properties of KNe drawn from compact object mergers. These differences may depend on the merger type (NS–NS versus NS–BH) and on the physical properties of the binary (mass ratio, spin periods, etc.).

Gompertz et al. (2018) found that for GRB 061201 a KN of the same brightness of AT 2017gfo could have been observed, but deep 3 $\sigma$ upper limits on this observation are likely to exclude 
the presence of a KN similar to AT 2017gfo.

The KNe event associated with GRB 130603B (Berger et al. 2013) and GRB 060614A (Yang et al. 2015) are 2 or 3 times brighter than the interpolated KN model fit.

Rossi et al. (2020), among 28 SGRBs, found seven of them associated with claimed KNe or with a shallow decay of the afterglow, which could be a signature of the KNe, with a known redshift. Out of these seven GRBs, three are present in our sample: GRB 060614A, GRB 070714B, and GRB 130603B.
GRB 070809 is associated with a KN, but with less secure redshift (Rossi et al. 2020).
GRBs 111117A and 100625A have a probability $> 1\%$ to be associated with KNe. However, given the lack of any other possible galaxy with similarly low chance association, these cases are more likely to be associated with KNe.
GRB 061201 has a luminosity smaller than $0.35$ of the luminosity of AT2017gfo. This is the possible reason why the KN has not been detected. GRB 140903A is 15 times brighter than the AT2017gfo, meaning that this burst could have masked the KN (Gompertz et al. 2018).

\begin{figure}[b]
\includegraphics[width=1.2\hsize,height=0.67\textwidth,angle=0,clip]{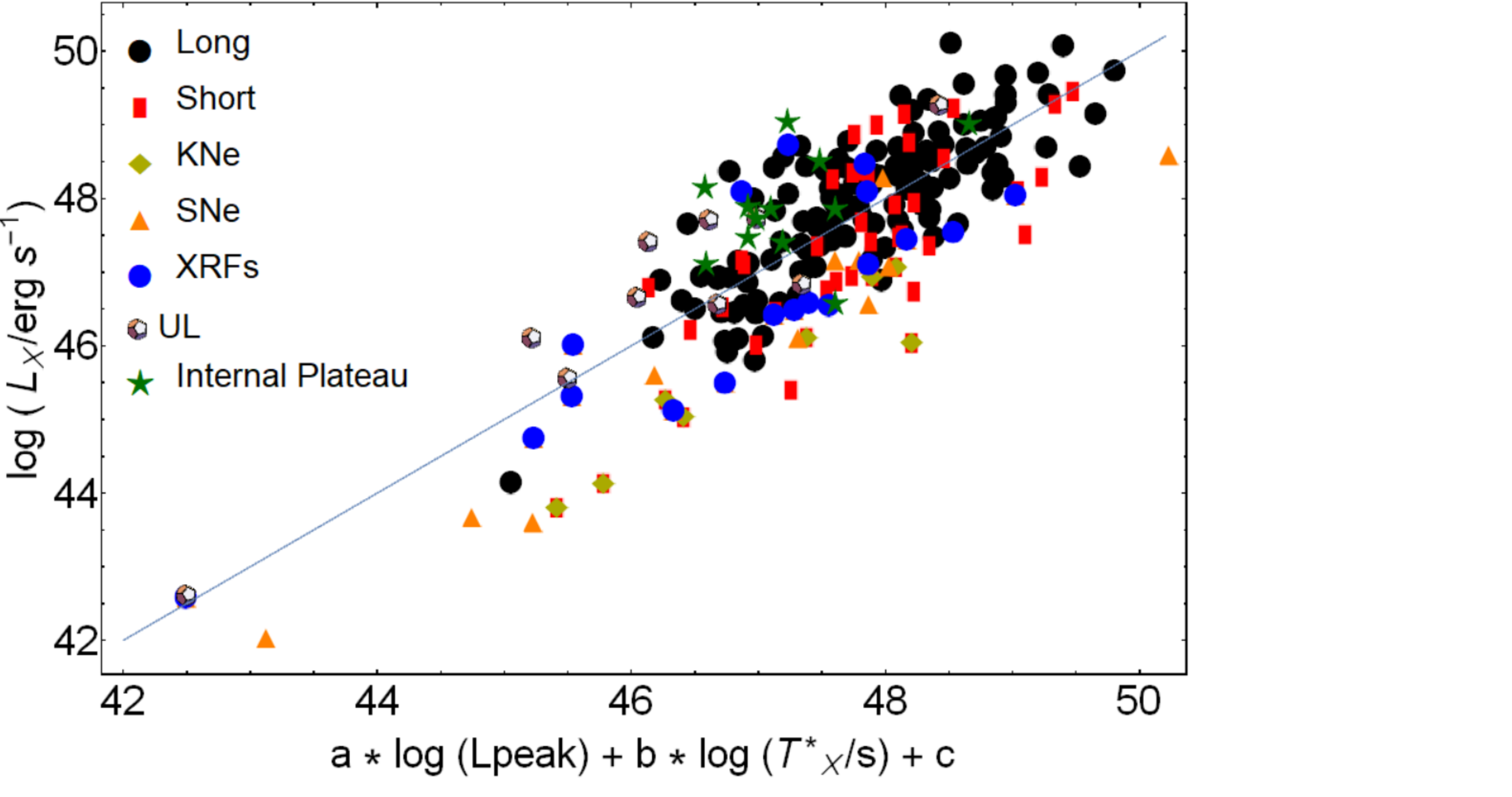}
\caption{The 2D projection of the $L_X- T^{*}_{X}-L_{peak}$ relation for the 222 GRBs of our sample, with a plane fitted including LGRBs (black circles), SGRBs (red rectangles), KN-SGRBs (dark yellow rhombuses), SN-LGRBs (orange triangles), XRFs (blue circles), ULGRBs (dodecahedrons), and GRBs with internal plateaus (green stars).}
\end{figure}
\label{fig2D}

\section{The 3D Relation for KN-SGRBs and the Other classes}\label{3D correlation}

\begin{figure}
\includegraphics[width=0.50\hsize,height=0.45\textwidth,angle=0,clip]{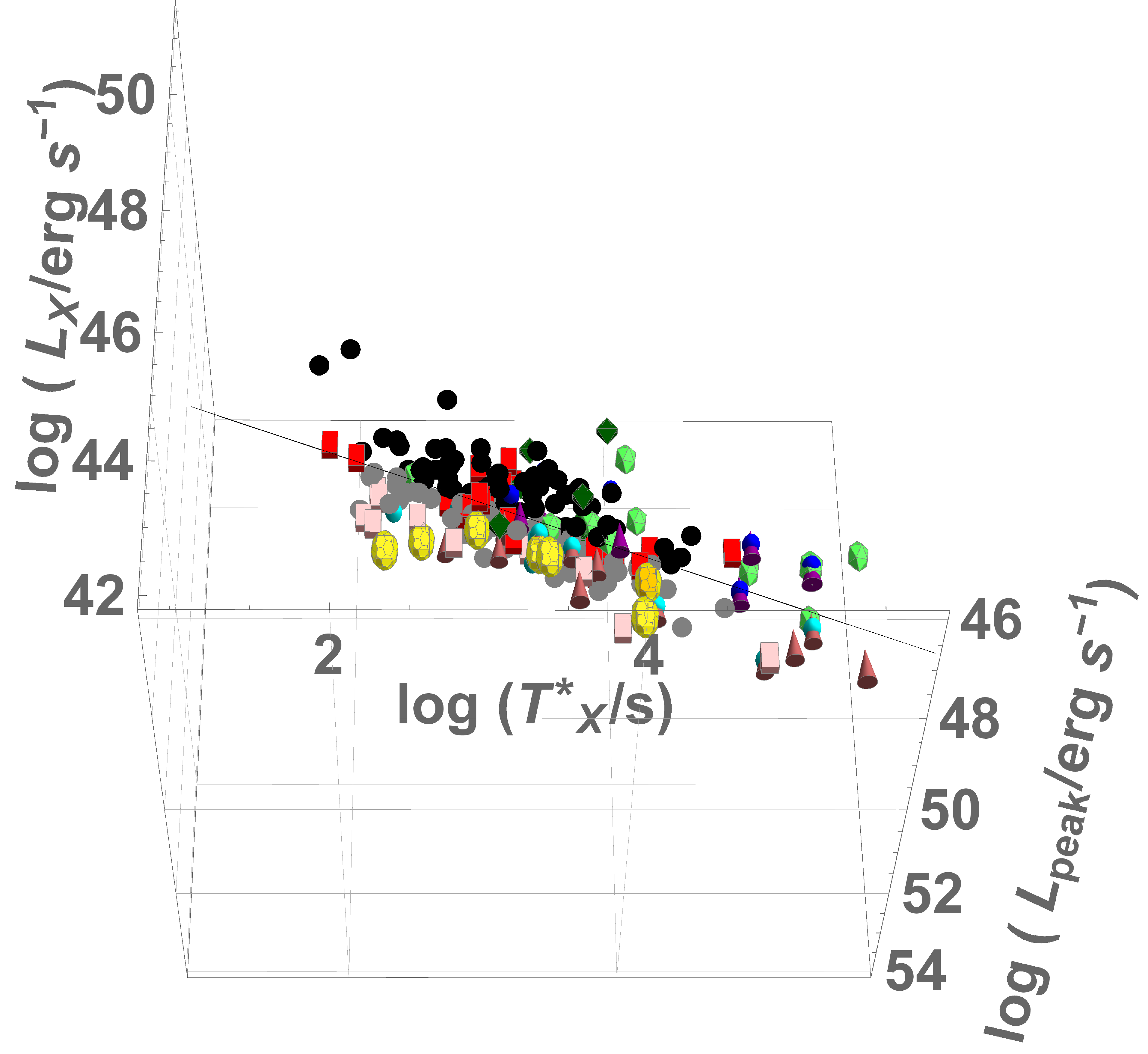}
\includegraphics[width=0.50\hsize,height=0.55\textwidth,angle=0,clip]{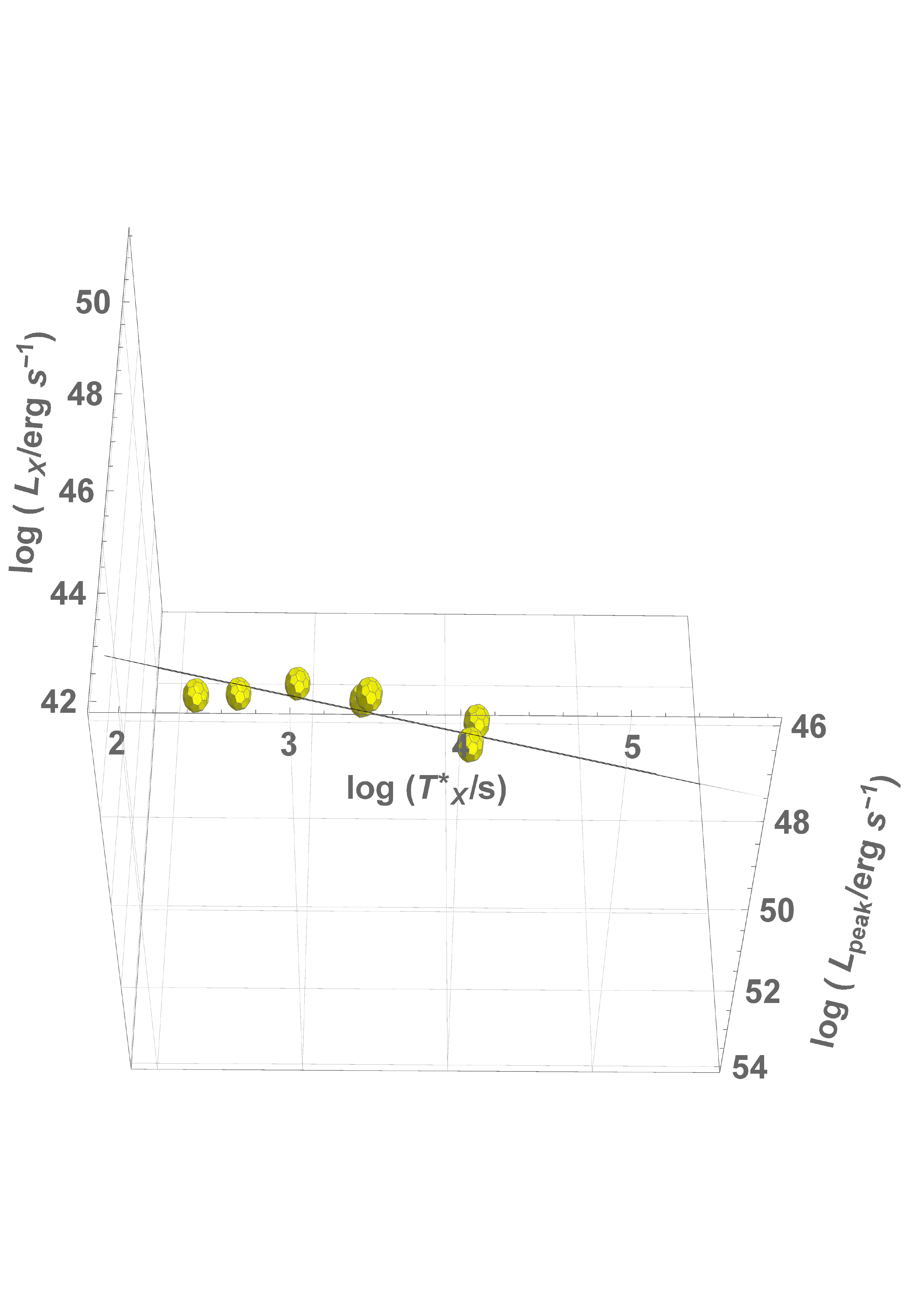}
\caption{Left panel: 222 GRBs in the $L_{X}-T^{*}_{X}-L_{peak}$ parameter space,  with a fitted plane including SN-LGRBs (purple cones), XRFs (blue spheres), SGRBs (red cuboids), LGRBs (black circles), ULGRBs (green dodecahedrons), KN-SGRBs (yellow truncated icosahedrons), and GRBs with internal plateau (dark green diamonds). Darker colors indicate GRBs above the plane, while lighter colors indicate GRBs below the plane. This figure shows the edge on projection.Right panel shows the same fitting, but with only the KN-SGRB.} 
\label{fig1}
\end{figure}

\begin{figure}
\includegraphics[width=0.333\hsize,height=0.35\textwidth,angle=0,clip]{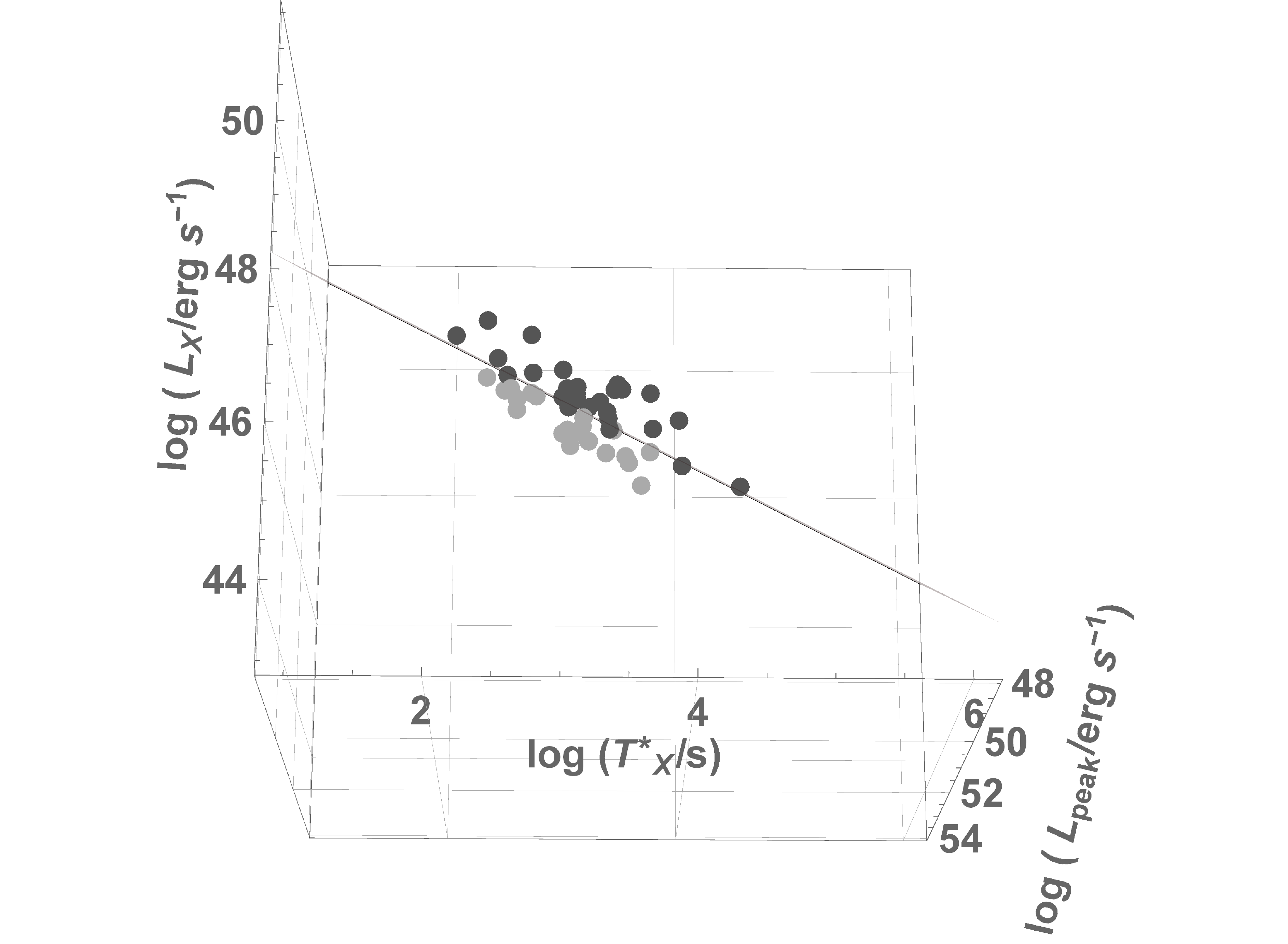}
\includegraphics[width=0.333\hsize,height=0.35\textwidth,angle=0,clip]{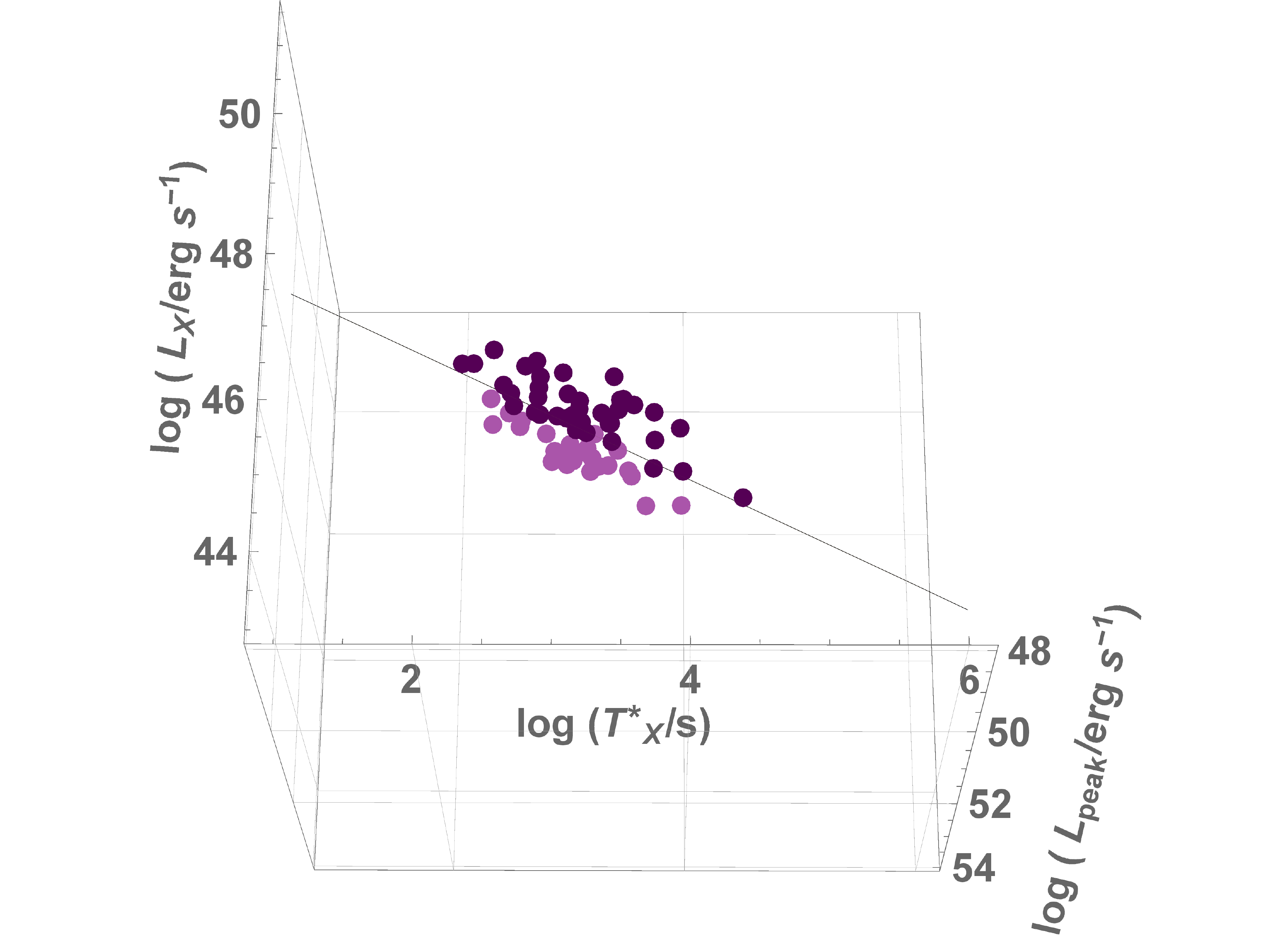}
\includegraphics[width=0.333\hsize,height=0.35\textwidth,angle=0,clip]{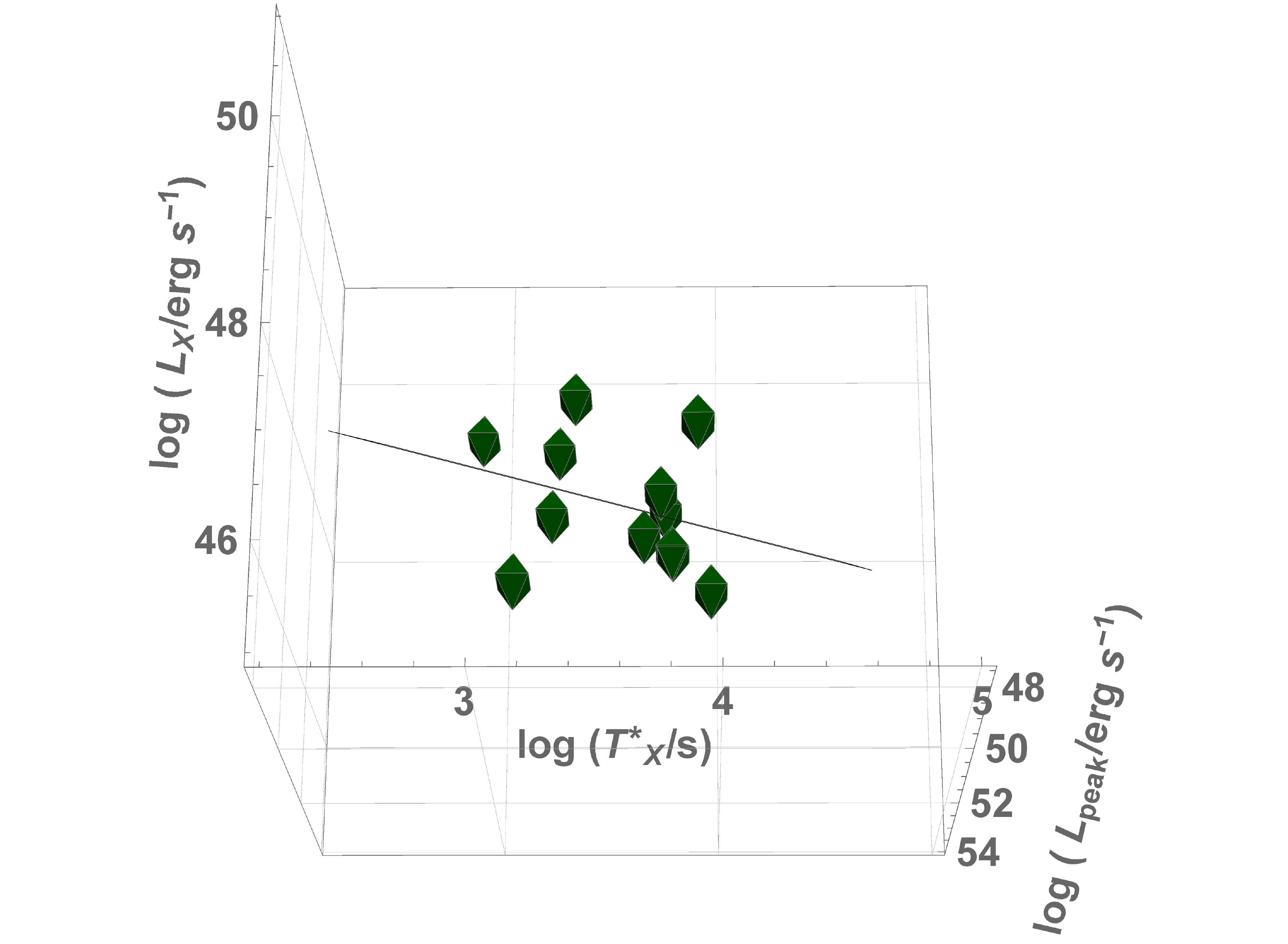}
\caption{The platinum, gold, and GRBs with internal plateau planes, respectively. Darker colors indicate GRBs above the plane, while lighter colors indicate GRBs below the plane. This figure shows the edge on projection. }
\label{fig1b}
\end{figure}

We perform all the fits using the D'Agostini (2005) method, which includes the intrinsic scatter, $\sigma_{int}$. Uncertainties are quoted in 1 $ \sigma$.
The fundamental plane relation is defined as follows:
\begin{equation}
\log L_X = C_o + a \times \log T^{*}_{X} + b \times \log L_{peak},
\end{equation}

\noindent where $C_o$ is the normalization, $a$ and $b$ are the best-fit parameters related to $\log T^{*}_{X}$ and $\log L_{peak}$, respectively.
In Figure 1 we show the 2D projection of the fundamental plane relation for all the 222 GRBs in our sample, considering the ones with internal plateau as well.
The left panel of Figure \ref{fig1} shows the same distribution in the $L_{X}-T^{*}_{X}- L_{peak}$ parameter space. 

The new gold sample is $44\%$ larger than the old gold one ($65$ versus $45$ GRBs). 
Here, $L_{peak}$ is computed giving preferences to the CPL rather than the PL. The best-fit parameters of the planes are shown in Table \ref{Table1}. The platinum sample yields a smaller $\sigma_{int}=0.34 \pm 0.04$ when we consider observed distributions (hereafter when we say the word “observed” we mean distributions or values for which selection biases are not accounted for), with a reduction of $\sigma_{int}$ of $12.8\%$ compared to the updated observed gold sample, and has a compatible intrinsic scatter in 1 $\sigma$ with the previous observed gold, $\sigma_{int}=0.32 \pm 0.04$ (Dainotti et al. 2017a), but with 2 more GRBs. The $\sigma_{int}$ of the updated observed gold Sample is comparable within $1$ $\sigma$ with respect to the previously observed gold.

The KN-SGRB sample has the smallest observed $\sigma_{int}=0.21 \pm 0.16$ with a reduction of $\sigma_{int}$ of $46.1 \%$ compared to the updated observed gold sample (right panel of Figure \ref{fig1}). The second smallest observed $\sigma_{int}=0.29 \pm 0.10$ is obtained by the SN-LGRB-ABC sample (bottom left panel of Figure 4), with a reduction of $27.5 \%$ compared to the updated observed gold sample. KN-SGRBs all fall below the plane of the total sample (see Figure 1 and the left panel of Figure \ref{fig1}), thus implying that the KNe fundamental plane is statistically different from the total sample. We note that the Type II observed sample gives us the largest intrinsic scatter: $\sigma_{int}=0.66 \pm 0.05$. We have computed the $\sigma_{int}$ of the whole sample as well, with and without the GRBs with internal plateaus, obtaining comparable results.

\begin{table}
\centering
\hskip -3.0cm    
\caption{The Observed Best-fit for the Plane Parameters (First Half of the Table) and Accounting for the Evolution, Indicated with the Subscript ``cor" (the Second Half of the Table), $\sigma_{int}$ and Number of GRBs for Each Category}
\begin{tabular}{| l | l | l | l | l | l | l | l | l | l | l |}
\hline
	Class & {\it a}& {\it b}& $C_o$ & $\sigma_{int}$ & {\it N}  &  {\it $a_{cor}$} & {\it $b_{cor}$} & $C_{ocor}$ & $\sigma_{intcor}$\\  \hline
	Gold & -0.82$\pm$0.13 & 0.55$\pm$0.11 &22.2$\pm$5.6 & 0.39$\pm$0.04 & 65   & -0.79$\pm$0.15  & 0.47$\pm$0.14 & 27.1$\pm$7.2 & 0.32$\pm$0.07  \\ \hline
	Platinum & -0.86$\pm$0.13 & 0.56$\pm$0.12 & 21.8$\pm$6.3 & 0.34$\pm$0.04 & 47 & -0.90$\pm$0.16 & 0.50$\pm$0.16 & 25.6$\pm$8.2  & 0.22$\pm$0.10 \\ \hline
	Long & -0.98$\pm$0.07 & 0.62$\pm$0.06& 19.1$\pm$3.1 & 0.43$\pm$0.03 & 129 & -1.05$\pm$0.09 & 0.65$\pm$0.09& 18.7$\pm$4.8 & 0.40$\pm$0.05 \\ \hline
	Short & -0.58$\pm$0.10 & 1.15$\pm$0.10 &-9.7$\pm$5.0 & 0.38$\pm$0.05 & 43  & -0.74$\pm$0.17 & 1.54$\pm$0.23 & -27.8$\pm$12.0 & 0.55$\pm$0.11\\ \hline
	SN-LGRB & -0.81$\pm$0.14 & 0.72$\pm$0.07 & 13.2$\pm$3.9 & 0.42$\pm$0.08 & 22   & -0.77$\pm$0.18 & 0.82$\pm$0.10 & 8.2$\pm$5.7 & 0.43$\pm$0.09\\ \hline
SN-LGRB-ABC & -1.16$\pm$0.16 &0.59$\pm$0.07 & 20.6$\pm$4.1 & 0.29$\pm$0.10 & 14  & -1.18$\pm$0.18 & 0.65$\pm$0.09 & 18.3$\pm$5.2 & 0.22$\pm$0.10  \\ \hline
	XRFs & -0.81$\pm$0.19 & 0.69$\pm$0.13 & 14.6$\pm$6.7 & 0.54$\pm$0.10 & 18   & -0.92$\pm$0.25 & 0.66$\pm$0.17 & 16.9$\pm$8.9 & 0.50$\pm$0.19  \\ \hline
	UL & -0.62$\pm$0.20 & 0.74$\pm$0.12 & 11.6$\pm$6.2 & 0.43$\pm$0.15 & 10   & -0.72$\pm$0.27 & 0.94$\pm$0.19 &2.9$\pm$9.8 & 0.51$\pm$0.23 \\ \hline 
	KN-SGRB & -0.83$\pm$0.22 & 0.80$\pm$0.25 & 8.5$\pm$12.9 &  0.21$\pm$0.16 & 8  & -1.09$\pm$0.20 & 1.03$\pm$0.27 & -1.5$\pm$13.3 & 0.24 $\pm$0.12 \\ \hline
	Type II & -1.15$\pm$0.08 & 0.28$\pm$0.05 & 37.2$\pm$ 2.6 & 0.66$\pm$0.05 &   167  & -1.14$\pm$0.09 & 0.28$\pm$0.06 & 37.3$\pm$3.0 & 
	0.66 $\pm$ 0.05 \\ \hline
	Int. plateau & -0.4$\pm$0.4 & 0.36$\pm$0.24 & 30.9$\pm$ 12.4 &  0.59$\pm$0.12 &  12  & -0.28$\pm$0.88 & 0.64$\pm$0.58 & 14.8$\pm$30.3 & 	0.55 $\pm$ 0.29 \\ \hline
	No int.plateau & -0.78$\pm$0.05 & 0.82$\pm$0.04 & 8.1$\pm$ 2.2 &  	0.50$\pm$0.03 &  210  & -0.93$\pm$0.08 & 0.88$\pm$0.08 & 6.2$\pm$4.0 & 	0.61 $\pm$ 0.04 \\ \hline
	Whole sample & -0.77$\pm$0.06 & 0.81$\pm$0.05 & 8.6$\pm$ 2.5 &  	0.52$\pm$0.03 &  222  & -0.91$\pm$0.08 & 0.87$\pm$0.08 & 6.9$\pm$4.1 & 	0.64 $\pm$ 0.04 \\ \hline
\end{tabular}

%Gold+alpha
\label{Table1}
\end{table}

Left and right panels of Figure \ref{fig1} show the fitted plane for all GRBs and KN-SGRB, while Figure \ref{fig1b} shows the platinum, gold and internal plateau classes in order of increasing $\sigma_{int}$. Upper and lower panels of Figure 4 show the ULGRB, the SN-LGRB, the SGRB, and the SN-LGRB-ABC samples. From left to right, both panels show decreasing observed $\sigma_{int}$. The intrinsic scatter of the SN-LGRB-ABC plane is smaller than that of the total SN-LGRB sample ($\sigma_{int,SN-LGRB-ABC}=0.29 \pm 0.10$ versus $\sigma_{int,SN-LGRB}= 0.42 \pm 0.08$). This confirms a previous study of the $L_X-T^{*}_{X}$ correlation on the same sample for which this class of GRBs yields a smaller $\sigma_{int}$ than the total SN-LGRB sample.

\begin{figure}
\includegraphics[width=0.5\hsize,height=0.4\textwidth,angle=0,clip]{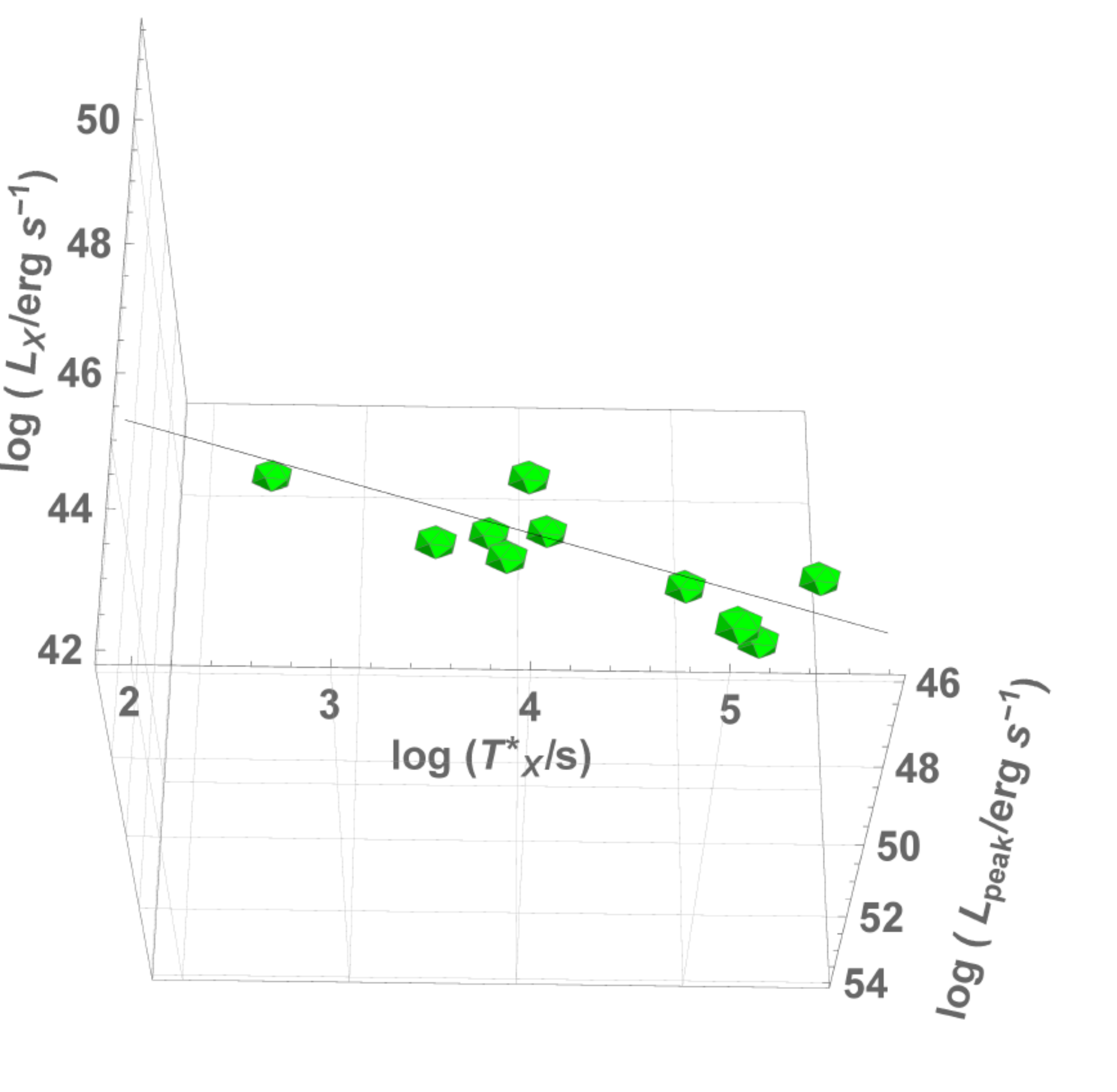}
\includegraphics[width=0.5\hsize,height=0.39\textwidth,angle=0,clip]{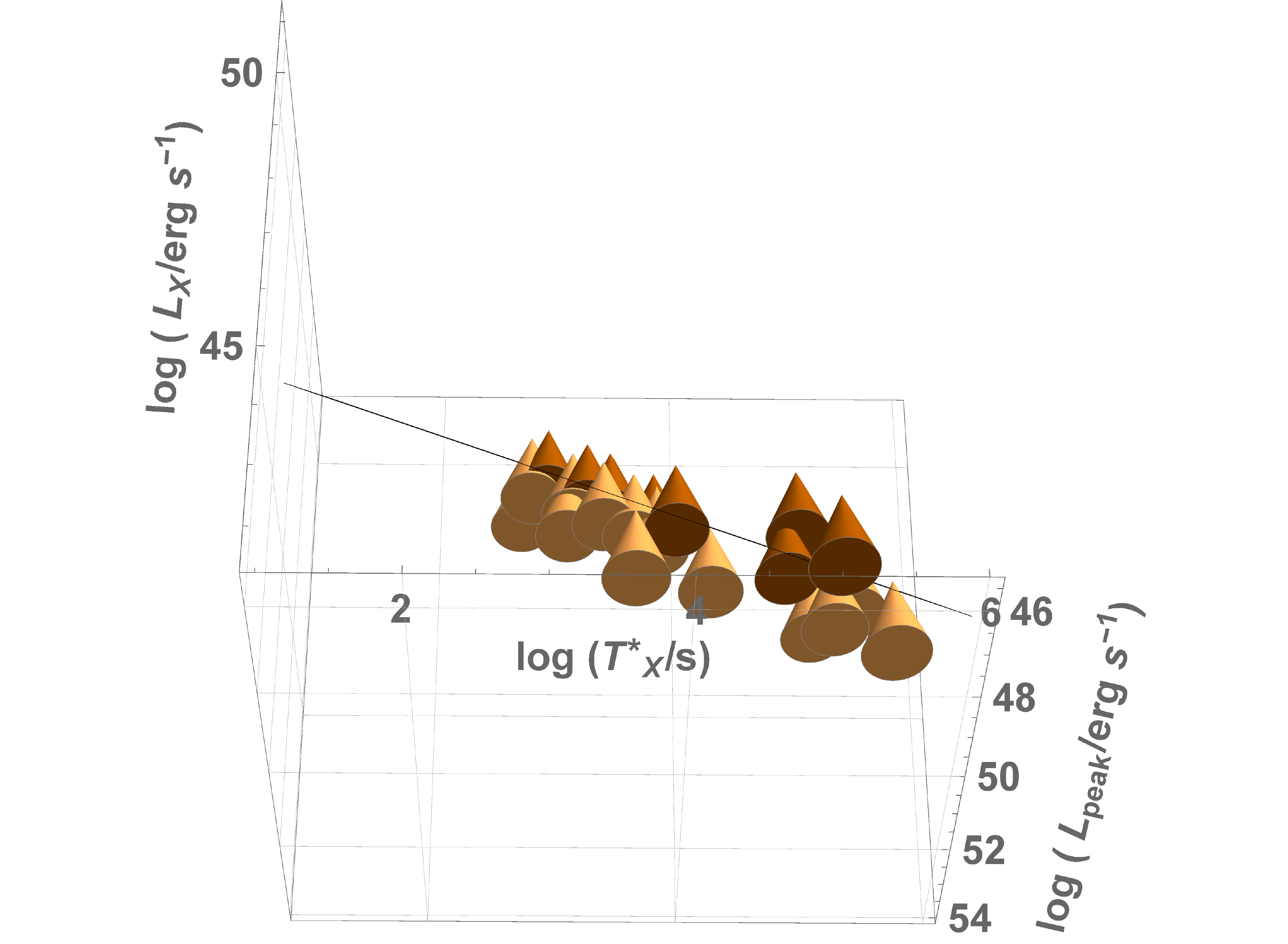}
\includegraphics[width=0.5\hsize,height=0.38\textwidth,angle=0,clip]{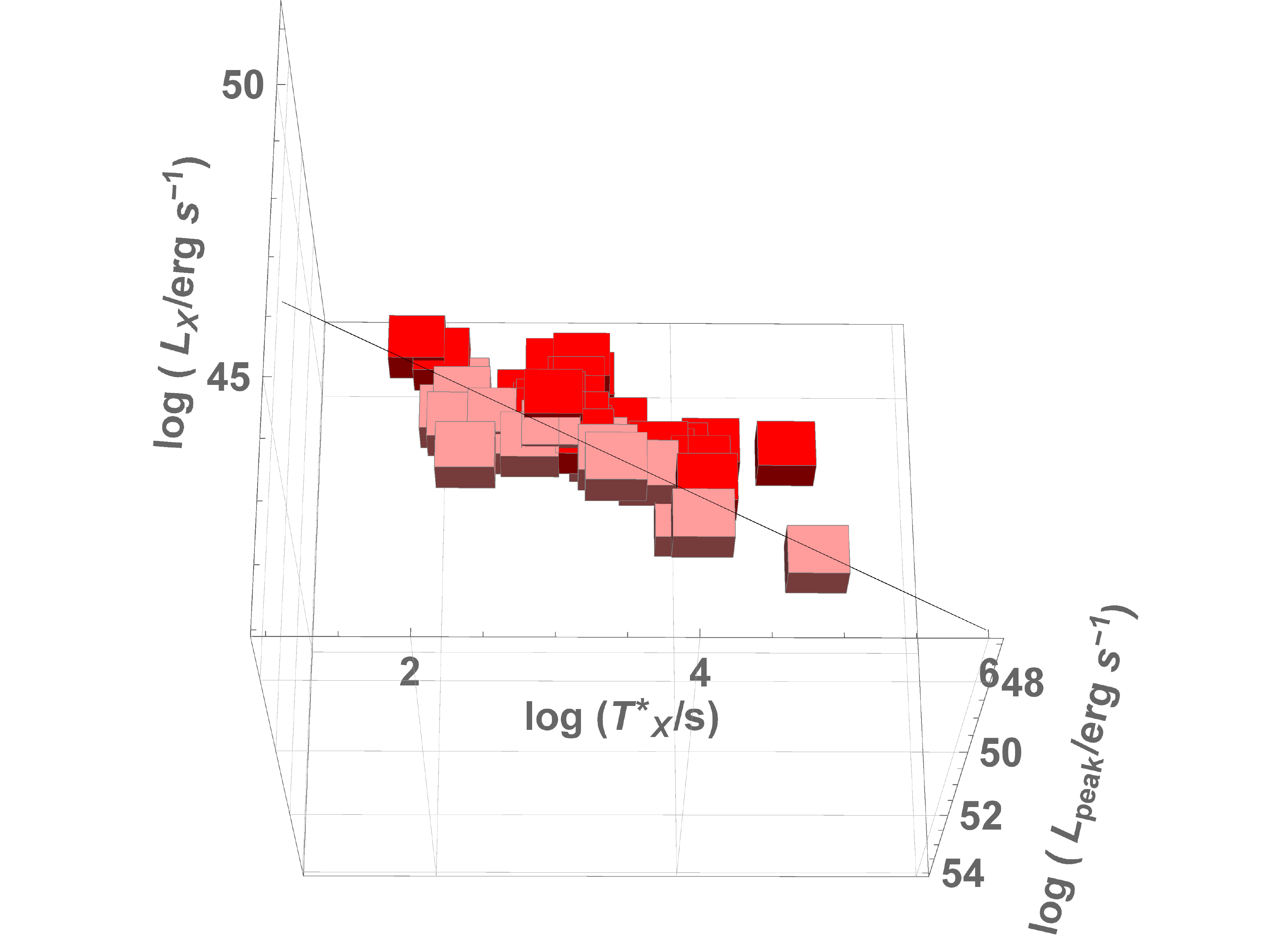}
\includegraphics[width=0.5\hsize,height=0.38\textwidth,angle=0,clip]{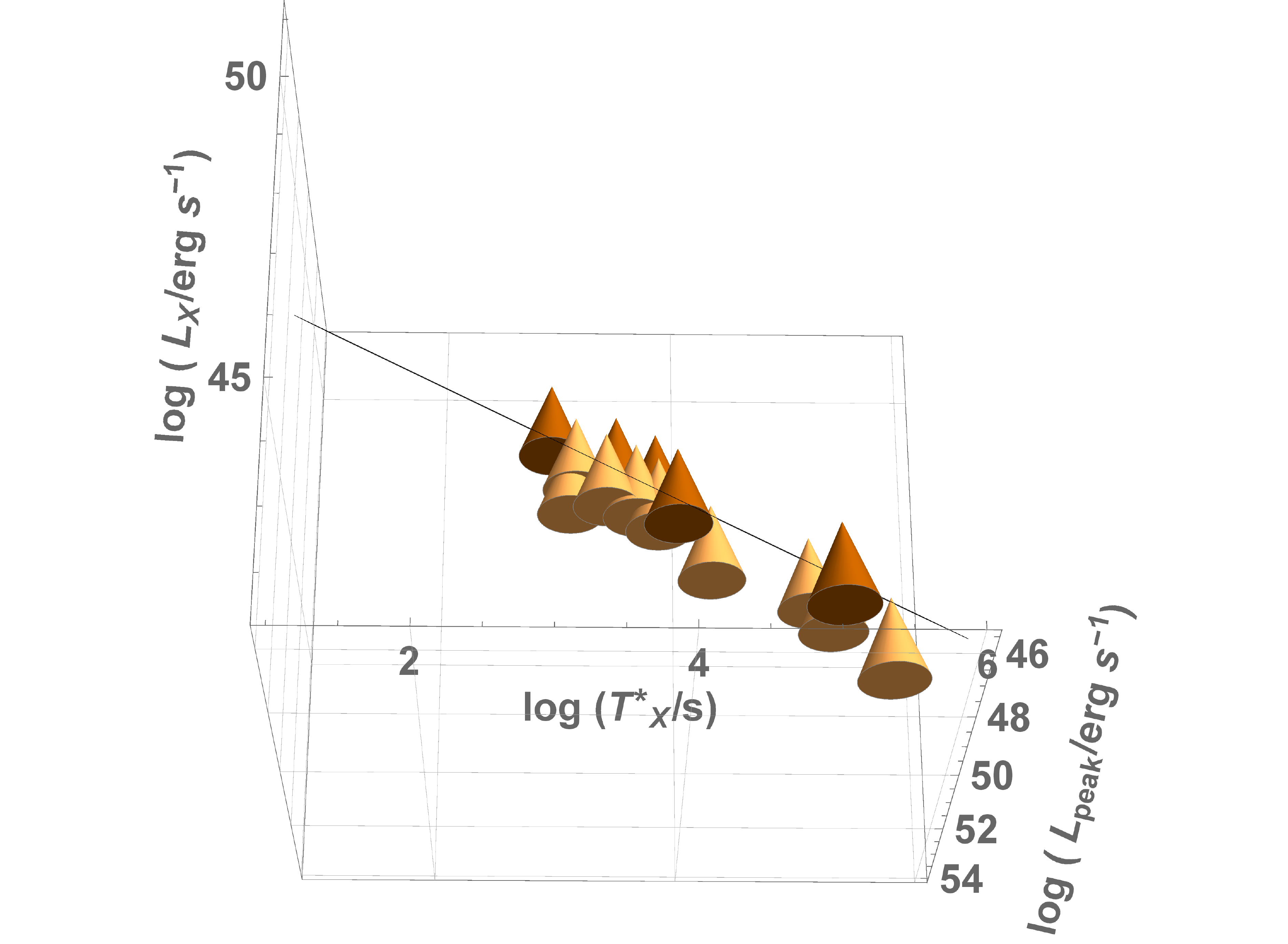}
\caption{The $L_X- T^{*}_{X}-L_{peak}$ relation, in order of decreasing $\sigma_{int}$, for ULGRB, SN-LGRB, SGRB, and SN-LGRB-ABC samples, respectively. Darker colors indicate GRBs above the plane, while lighter colors indicate GRBs below the plane. This figure shows the edge on projection.}
\end{figure}
\label{fig2}
\noindent 

All categories have high values of {\it $R^2$} and {\it $R_{adj}^2$} \footnote{{\it $R_{adj}^2$} is a version of the coefficient of determination, {\it $R^2$}, adjusted for the number of parameters in the model.}.
Particularly, the highest $R_{adj}^2$ are for SN-LGRB-ABC, SN-LGRB, KN-SGRB, ULGRB, and SGRB samples=(0.97, 0.93, 0.90, 0.88, 0.87), and the highest $R^{2}$ are for SN-LGRB-ABC, SN-LGRB, KN-SGRB, ULGRB, and SGRB samples=(0.97, 0.93, 0.92, 0.90, 0.88) for the observed distributions. A very low $p$-value (the probability of the same sample occurring by chance) yields for all categories: P (gold, platinum, long, short, SN-LGRB, SN-LGRB-ABC, XRF, ULGRB, KN-SGRB)=$(7 \times 10^{-14}, 2 \times 10^{-13}, 3 \times 10^{-36}, 8 \times 10^{-20},2 \times 10^{-12},1 \times 10^{-9}, 3 \times 10^{-10},1 \times 10^{-5},1 \times 10^{-3})$.

We check the compatibility of the gold fundamental plane observed parameters ($a$, $b$, $C_0$) with the other classes. The platinum, long, SN-LGRB, ULGRB, KN-SGRB, XRF and internal plateau parameters are all compatible in 1 $\sigma$ with respect to the gold parameters. For the sample for which no internal plateau is included, $a$ is compatible in 1 $\sigma$, $b$ and $C_0$ in 2 $\sigma$. For SN-LGRB-ABC, $a$ is compatible in 2 $\sigma$, $b$ and $C_0$ in 1 $\sigma$. For the SGRB sample there is compatibility in $2$ $\sigma$ for $a$, in $3$ $\sigma$ for $b$ and in $3.1$ $\sigma$ for $C_0$. For the Type II GRBs there is a 2 $\sigma$ compatibility for $a$, $b$, and $C_0$. We note that the KN-SGRB plane $a$ and $b$ parameters are compatible within 1 $\sigma$ with respect to the SGRB ones, while the $C_0$ parameter is compatible within 2 $\sigma$, as expected from their physical origin. The differences in the fitting parameters of the fundamental plane relation could suggest different physical mechanisms or the same mechanisms, but with 
different environments, thus making these planes useful to test theoretical models (Srinivasaragavan et al. 2020). This feature is additionally highlighted by the $z$-score test for comparing two samples, computed as follows:

\begin{equation}
z=\frac{<x_1>-<x_2>}{\sqrt{\frac{\sigma_1^{2}}{N_1}-\frac{\sigma_2^{2}}{N_2}}},
\end{equation}

where $<x_i>$ and $N_i$ are the means and the sizes of the samples. We here stress that the $ z$-score in its formulation includes the number of GRBs in each subsample, thus each of them is weighted according to its size. We compute the $ z$-score for all classes with respect to the gold sample, then we use it to compute %and 
the probability, $P$, for each sample of being statistically compatible with the gold one, see Table \ref{Table2}. The KN-SGRB plane has the highest $ z$-score=10.18, corresponding to $P<10^{-4}$, and the two samples are drawn from the same population, thus showing that this class is a clear outlier together with the SN-LGRB, the SN-LGRB-ABC, and the SGRB classes.
 We refer to "outliers" as the classes that have at least one of the samples observed or corrected above $z$-score $\ge \mid 4 \mid$. 
This result is a hint that these categories can be produced by a distinct physical mechanism: KNe may be related to SGRBs and hence come from a different progenitor compared to the LGRBs considered in the gold and platinum samples. 

\begin{figure}
\includegraphics[width=1\hsize,height=0.6\textwidth,angle=0,clip]{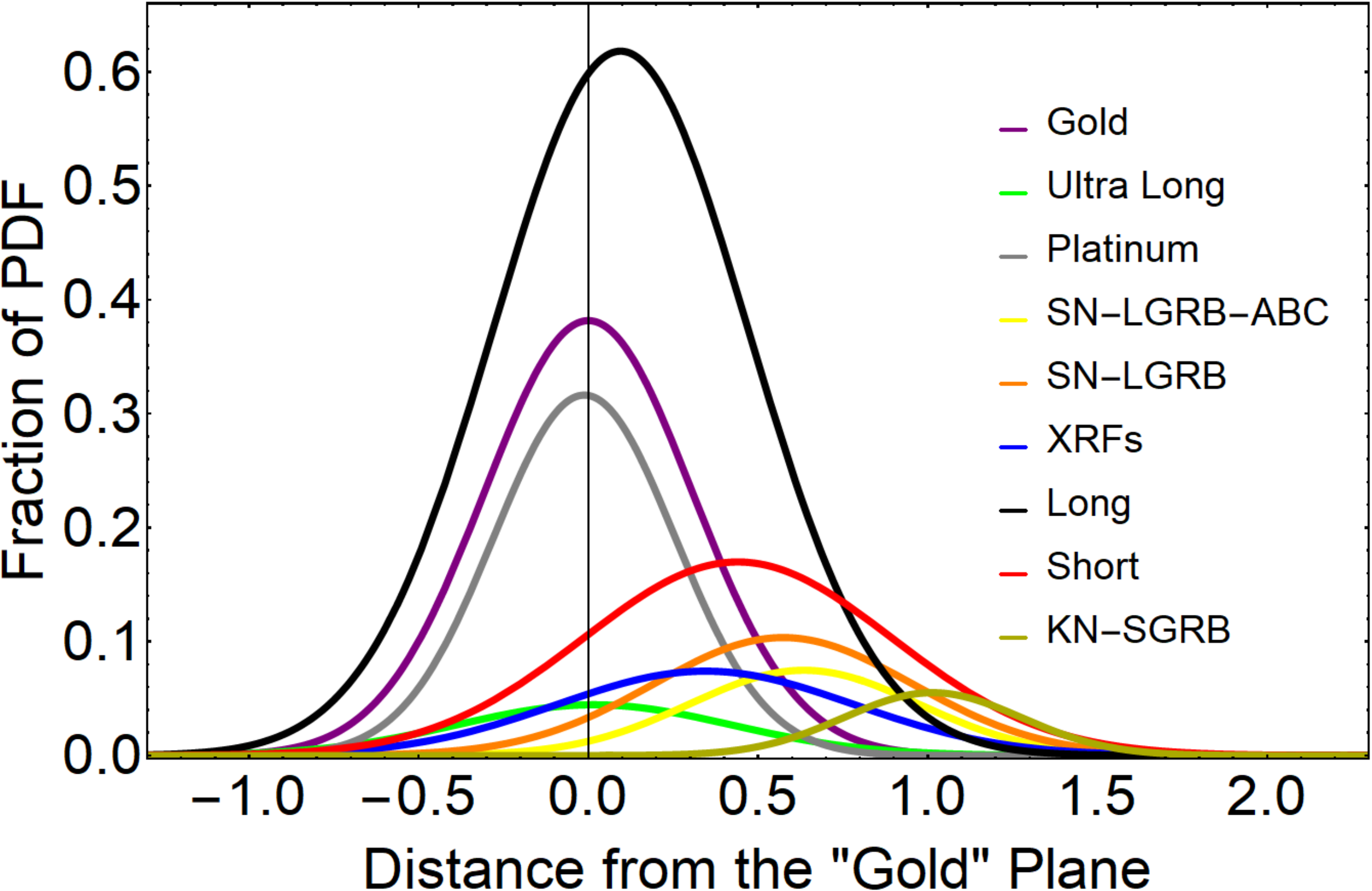}
\includegraphics[width=1\hsize,height=0.6\textwidth,angle=0,clip]{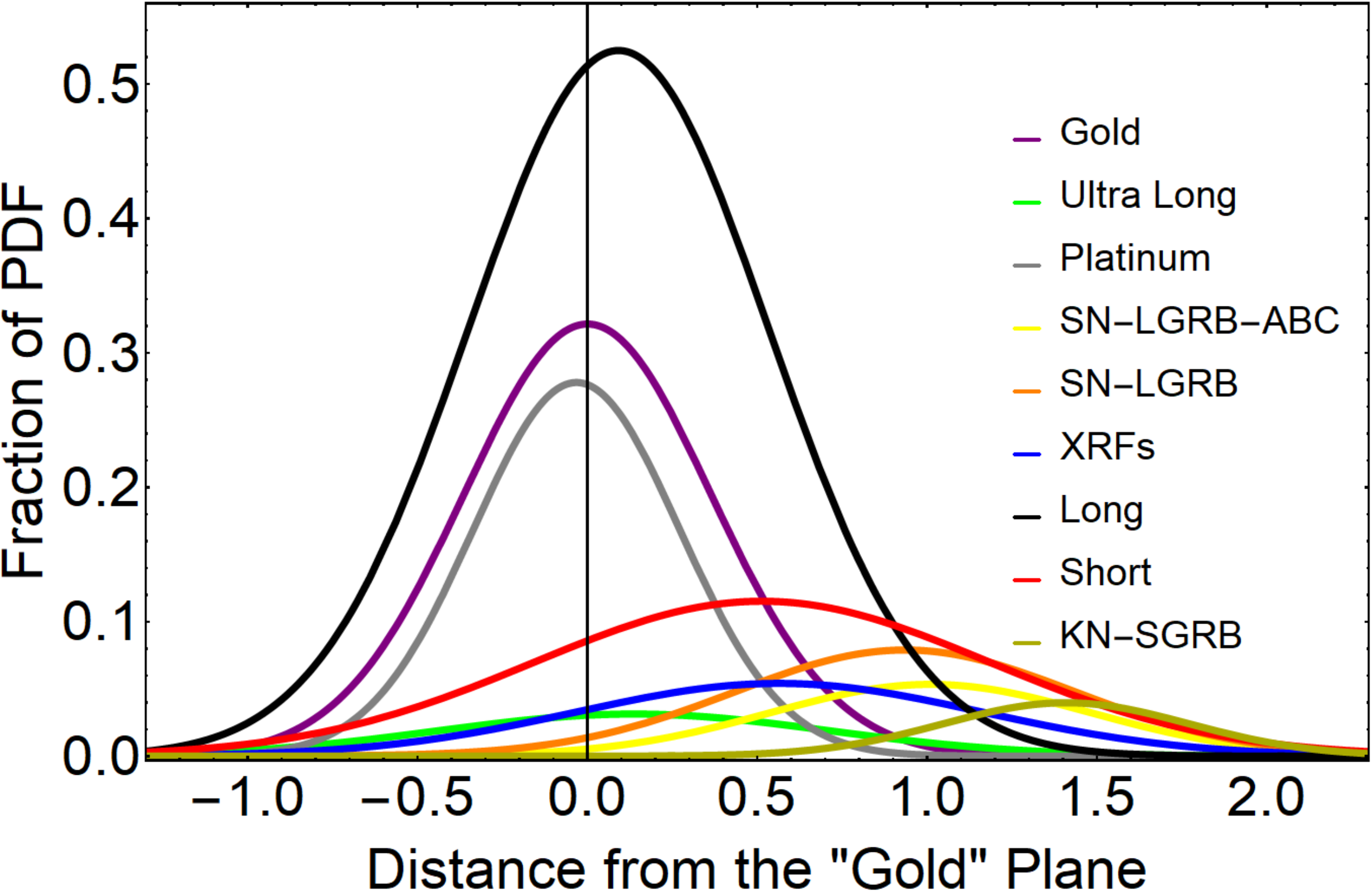}
\caption{Gaussian fits to the histogram of the distance distribution from the gold fundamental plane for all classes. The upper panel shows the fits without the correction for selection effects and redshift evolution, while the lower panel accounts for them. A line perpendicular to $x=0$ is shown as the reference of the gold sample.}
\label{fig3}
\end{figure}

\begin{figure}
\includegraphics[width=1.0\hsize,height=0.55\textwidth,angle=0,clip]{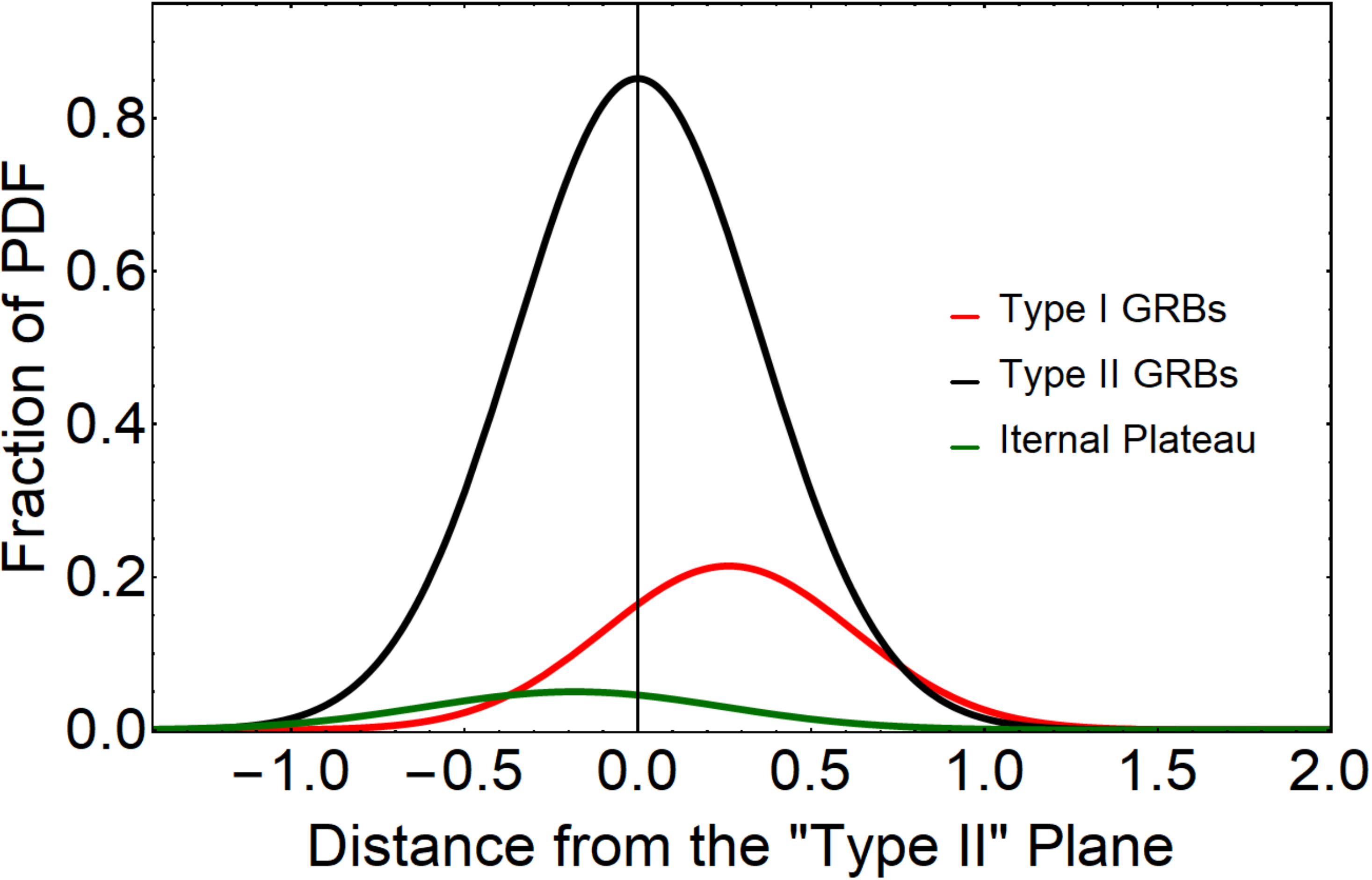}
\includegraphics[width=1.0\hsize,height=0.55\textwidth,angle=0,clip]{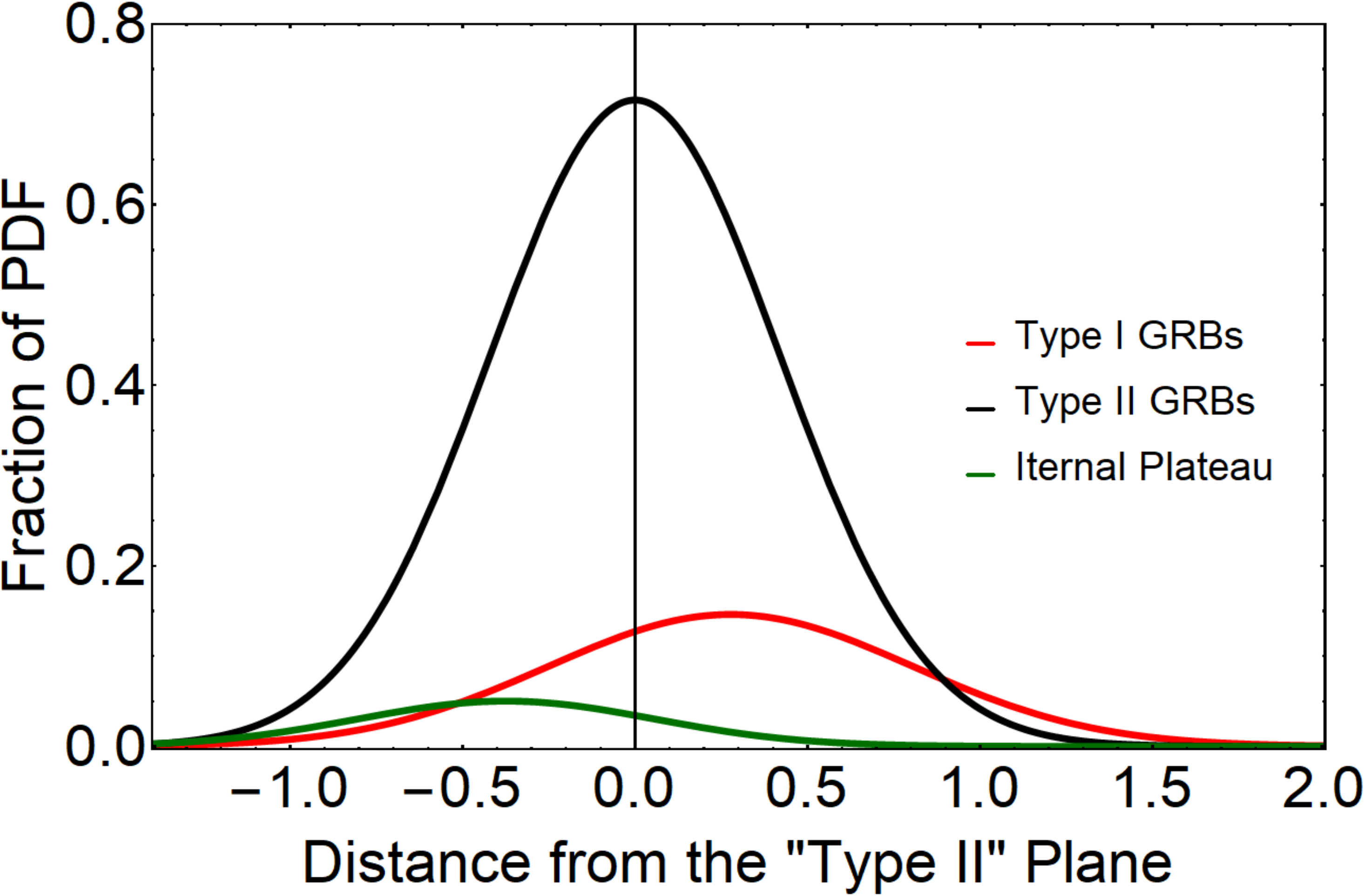}
\caption{Gaussian fits to the histogram of the distance distribution from the Type II fundamental plane for Type I, Type II, and internal plateau classes. A line perpendicular to $x=0$ is shown as the reference of the Type II sample. In the upper panel, the fit does not take into account the correction for selection biases and evolutionary effects, while in the lower panel it does.}
\label{fig4}
\end{figure}

\begin{figure}
\includegraphics[width=0.5\hsize,height=0.29\textwidth,angle=0,clip]{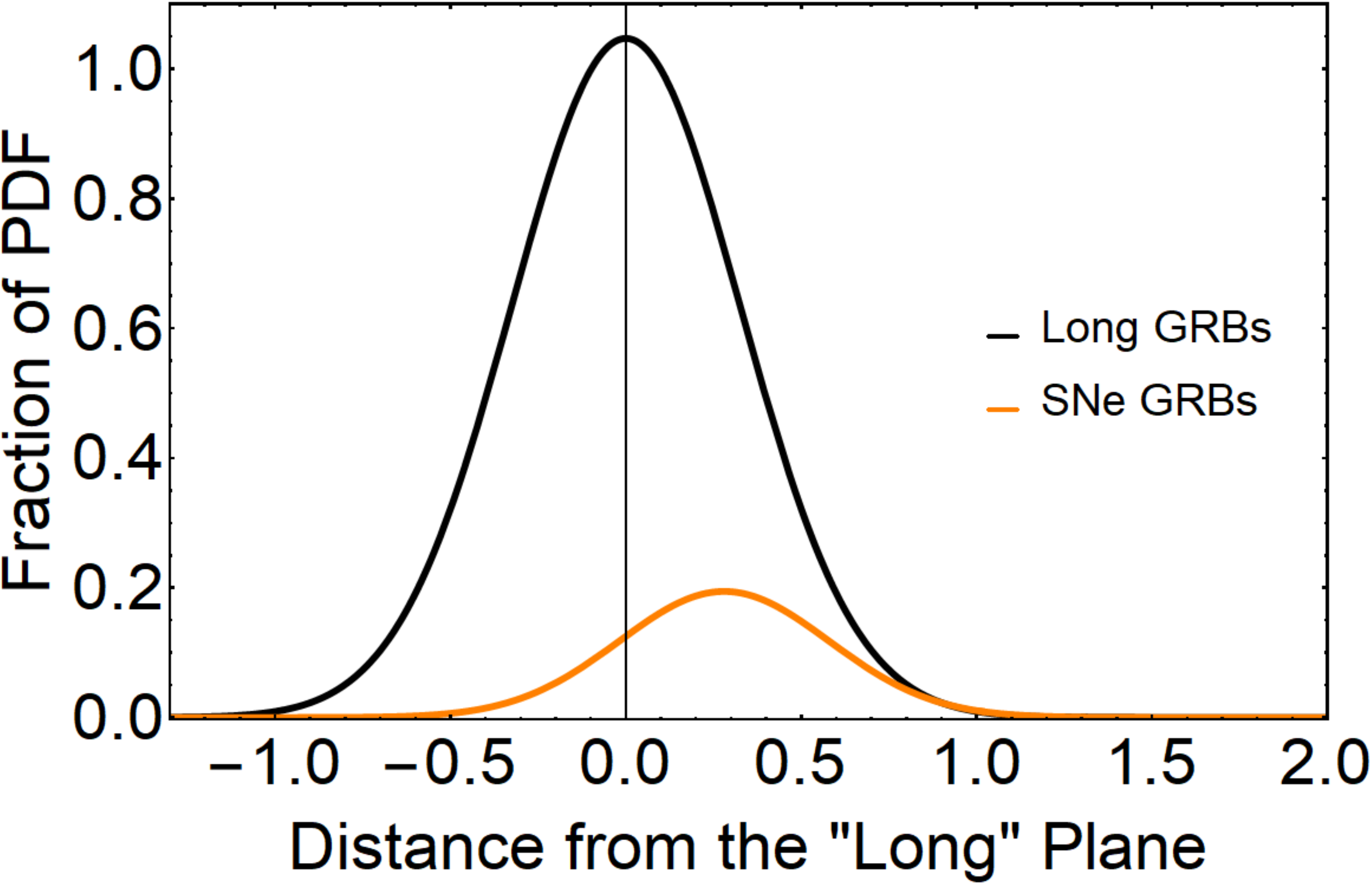}
\includegraphics[width=0.5\hsize,height=0.29\textwidth,angle=0,clip]{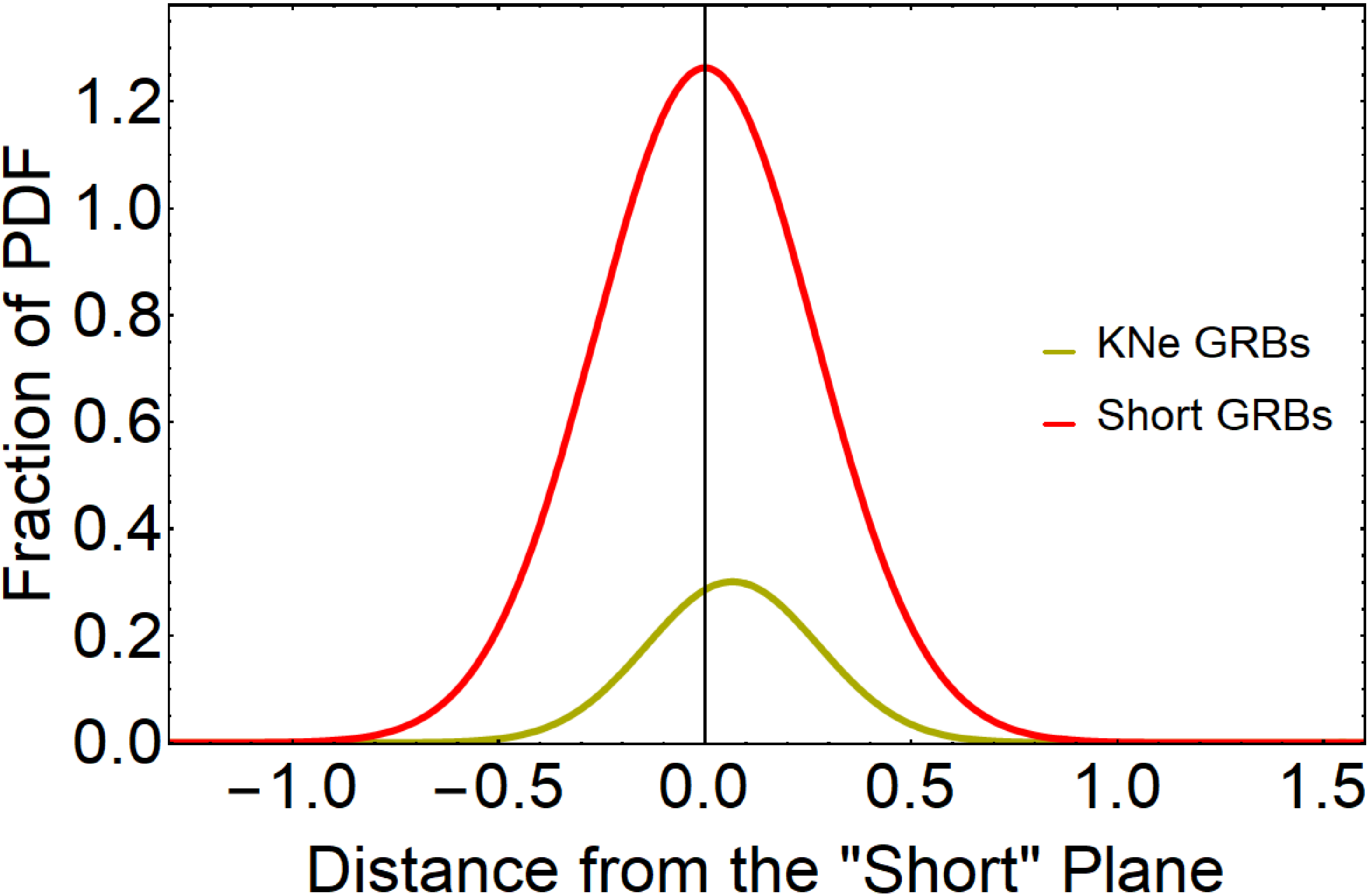}
\includegraphics[width=0.5\hsize,height=0.29\textwidth,angle=0,clip]{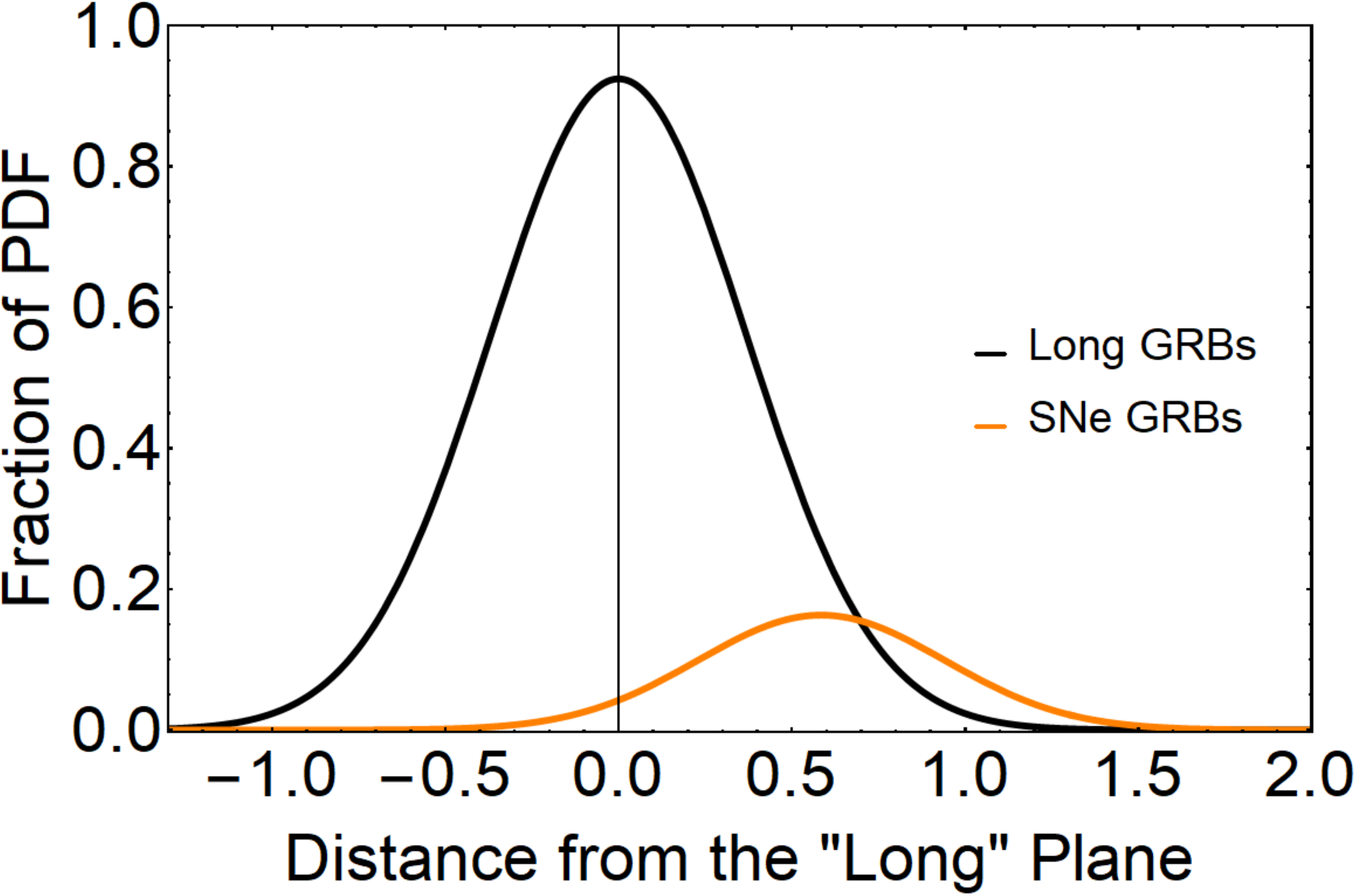}
\includegraphics[width=0.5\hsize,height=0.29\textwidth,angle=0,clip]{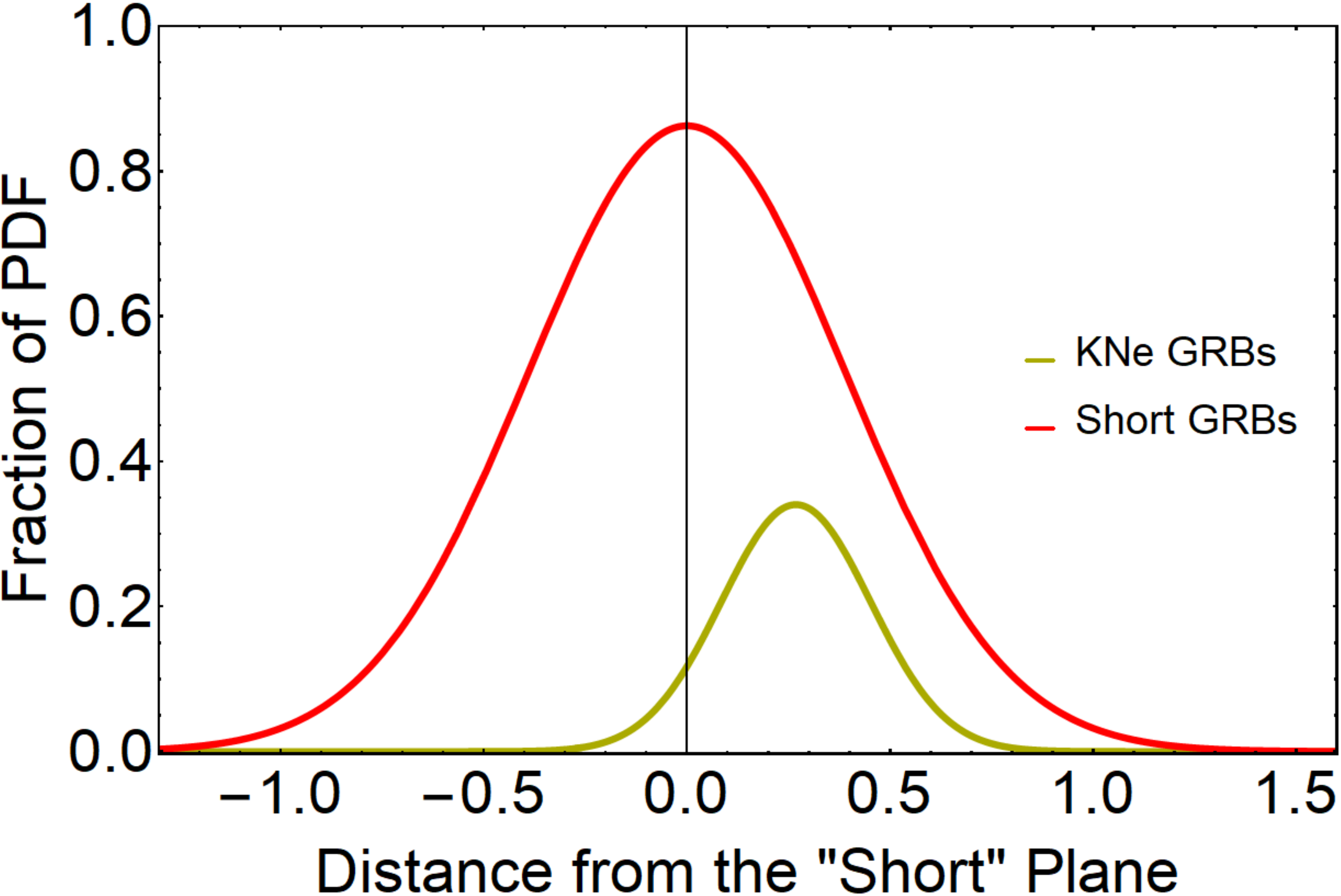}
\caption{Left panels: Gaussian fits to the histogram of the distance distribution from the long fundamental plane for the LGRB and SN-LGRB classes, without considering evolutionary effects (upper panel) and considering them (lower panel). Right panels: the same as the left panels with SGRB and KN-SGRB samples from the short fundamental plane.}
\label{fig5}
\end{figure}

The clear difference between the observed KN-SGRB plane and the others is evident in Figure \ref{fig3}, where the Gaussian distributions of the geometric distance from the gold fundamental plane are shown for each category. The Gaussian fits in Figures \ref{fig3}-\ref{fig5} represent fractional probability distribution functions (PDFs) obtained so that the size of each class with respect to the whole sample is taken into account. For instance, the fraction of the PDF related to the gold sample has been obtained by simply multiplying the PDF for 65/222, which is the size of the gold sample divided by the size of the total sample. In the upper panels of Figure \ref{fig3}-\ref{fig5}, the selection effects have not been considered, while in the lower ones they have been taken into account (see \S \ref{3D correlation with evolution}). The center of the distributions of KN-SGRBs, SGRBs, and SN-LGRBs are the furthest from the gold fundamental plane. The difference between the SN-LGRB and the gold samples have already been pointed out in Dainotti et al. (2017a), where a high $z$-score among those two classes have been found to be equal to -5.8. This strengthens the possibility that the distance to the gold fundamental plane is a relevant discriminant between categories. The {\it z}-score for the observed ULGRBs is very low ({\it z}-score=0.12), confirming that ULGRBs and LGRBs may belong to the same population (Zhang et al. 2014). This conclusion has been predicted in Dainotti et al. (2017a), where only two ULGRBs were considered.

\begin{table}
\begin{center}
\caption{Table of {\it z}-scores for All classes, with the Number of GRBs in Each Sample and the Probability that the Gold Fundamental Plane and the Other Planes Are drawn by the Same Distribution. }
\begin{tabular}{ |c|c|c|c|c|c|c|c| } 
 \hline

 Class & $z$-score & $N$ & Probability &  $z$-score cor & Probability cor\\ \hline

 Gold & 0.00 & 65 & 1.00 & 0.00 & 1.00 \\ 
 Platinum & -0.22 & 47 & 0.83 & -0.51 & 0.61 \\ 
 Long & 1.89 & 129 & 0.06  & 1.54 & 0.12\\
 SN-LGRB & 6.39 & 22 & $\le 10^{-4}$  & 8.07 & $\le 10^{-4}$\\ 
 SN-LGRB-ABC  & 6.51 & 14 & $\le 10^{-4}$  & 7.55 & $\le 10^{-4}$\\ 
 XRFs & 3.15 & 18 & 0.002  & 3.80 & 0.0001\\ 
 SGRBs & 5.57 & 43 & $\le 10^{-4}$  & 4.60 & $\le 10^{-4}$\\ 
 Ultra Long & 0.12 & 10 & 0.90  & 0.73 & 0.47\\ 
 KN-SGRB & 10.18 & 8 & $\le 10^{-4}$ & 10.39 & $\le 10^{-4}$\\ 
 \hline
\end{tabular}
\item \textbf{Note.} On the right side of the table we show {\it z}-scores without evolution, while on the left side the ones with evolution (cor).
\label{Table2}
\end{center}
\end{table}

%Type I and Internal Plateau vs Type II 

\begin{table}
\begin{center}
\caption{Table of {\it z}-scores for Type II, Type I and internal Plateau Classes with Respect to the Type II One, Including the Information of the Number of GRBs in Each Sample and the Probability that the Type II Fundamental Plane and the other Planes are Drawn by the Same Distribution.}
\begin{tabular}{ |c|c|c|c|c|c|c|c| } 
 \hline

 Class & z-score & N & Probability &  z-score cor & Probability cor\\ \hline

 Type II & 0.00 & 167 & 1.00 & 0.00 & 1.00 \\ 
 Type I & 4.28 & 43 & $\le 10^{-4}$ & 3.19 & 0.001 \\ 
 Internal plateau & -1.43 & 12 & 0.15  & -2.92 & 0.004\\

 \hline
\end{tabular}
\item \textbf{Note.} On the right side of the table we show {\it z}-scores without evolution, while on the left side the ones with evolution (indicated with the subscript cor).
\label{Table5}
\end{center}
\end{table}

We have computed the {\it z}-score for the Type I and internal plateau subsamples with respect to Type II. The results are shown in Table \ref{Table5}. We note a very low {\it z}-score between the internal plateau and Type II observed samples, as expected because all the internal plateau GRBs in our sample belong to the Type II GRBs. We have then computed the {\it z}-scores for distances from the long fundamental plane for the LGRBs and SN-LGRBs, see the left panels of Figure \ref{fig5}. The upper panel of Figure \ref{fig5} takes into account the correction for evolution, while the lower panel does not consider the evolution. Analogously, we have then computed the {\it z}-scores for the distances from the short fundamental plane for the SGRBs and KN-SGRBs, as shown in Table \ref{Table6} and in the right panels of Figure \ref{fig5}, where the upper panel shows results not corrected for the redshift evolution, while the lower panel includes corrections for the redshift evolution. Here, we note a low value of the {\it z}-score for the KN-SGRBs versus SGRBs ($z$-score=0.67 and $z$-score=1.91 without and with evolution, respectively), since KN-SGRB is a subsample of the SGRB one. In Figure \ref{fig4} the reference point is the Type II fundamental plane, while in Figure \ref{fig5} it is the LGRB plane for the left panels and the SGRB plane for the right ones.

In Figure 8 we focus our attention on the observed SGRB and KN-SGRB samples. Here, we note that even if the KN-SGRBs are all part of the SGRB sample a clear clustering of these GRBs is visible: the KN-SGRBs are positioned at lower $L_{peak}$ and $L_{X}$ values and they all lie below the short fundamental plane, thus showing that the KN-SGRB class has observational features different from the SGRB ones. However, a further investigation related to selection biases needs to be performed in order to verify if this clustering is intrinsically due to a physical mechanism.

In Figure 9 we show the paired histograms of the distances of the KN-SGRBs and the SGRBs from the SGRB plane (left panels) and the distances from the LGRB plane of the LGRBs and SN-LGRBs (right panels) taking into account the evolution (upper panels) and not considering the evolution (lower panels).

\section{The 3D Relation considering evolution} \label{3D correlation with evolution}

In a series of papers we have discussed the role of selection biases and redshift evolution for the $L_X-T^{*}_{X}$ (Dainotti et al. 2013) and the $L_X-L_{peak}$ relations (Dainotti et al. 2015b, 2017b), where we have discussed how selection biases and evolutionary effects change if we consider only the LGRB sample. Each variable, $L_X$, $T^{*}_{X}$ and $L_{peak}$, undergoes selection biases due to instrumental thresholds and redshift evolution.
To overcome this problem we use the EP method, which
employs a modification of the Kendall $\tau$ test to compute the statistical dependence among variables. $\tau$ is defined as 
\begin{equation}
\tau = {{\sum_{i}{(\mathcal{R}_i-\mathcal{E}_i)}} \over {\sqrt{\sum_i{\mathcal{V}_i}}}}
\label{tau}
\end{equation}
where $R_i$ is the rank, $\mathcal{E}_i=(1/2)(i+1)$ is the expectation value, and $\mathcal{V}_i=(1/12)(i^{2}+1)$ is the variance. 
The rank $R_i$ for each data point will be determined from its position in the “associated sets,” which include all objects that could have been detected given the observational limits as shown in (Dainotti et al. 2013, 2015b, 2017b)  and in Petrosian et al. (2015).
In this case, these limits are the luminosities and times. 
First, the luminosity and time evolutions, namely their dependence on the redshift for $L_{X}$, $T^{*}_{X}$ and $L_{peak}$ will be computed. This procedure is the same for all of these variables. To derive the $L_{X}$ and $L_{peak}$ evolution, the flux limit, $f_{lim}$, at the end of the plateau phase shall be determined. Then, the minimum luminosity will be computed, namely the luminosity that would allow the object to still be visible with a given redshift: $L_{min}(z_i)=4\pi D_L(z_i)^2f_{lim} K$. Similarly, $T^{*}_{X,lim}=T_{X,lim}/(1+z)$, where $T_{X, lim}$ is the minimum end time of the plateau for a given observed sample and energy band. The associated set for a GRB at a given $z_i$ contains all objects that have luminosity $L_j \ge L_{min}$ and redshift $z_j \leq z_{i}$. The objects in the sample and in the associated sets are indicated with {\it i} and {\it j}, respectively. 
The EP procedure requires conservative choices for these limiting values, such that the samples used are at least $90\%$ of the original ones. Therefore, this method enables us to remove biases without substantially reducing the samples, and its reliability has been already verified with Monte Carlo simulations (Dainotti et al. 2013).
Since the evolution of the parameters is determined for a smaller sample with the EP method with less precision, and since the evolutionary effects are compatible within $2$ $\sigma$ between the LGRBs and the total samples we used as evolutionary functions the ones quoted in Dainotti et al. (2017b). The results are tabulated in the last four columns of Table \ref{Table1}.

All the samples present a 1 $\sigma$ compatibility for all the plane parameters compared to the ones without evolution (the only exceptions are the $b$ and $C_{0}$ parameters for the SGRB sample, which are compatible with the ones without evolution within 2 $\sigma$). 
After the redshift evolution and selection biases are removed, the platinum and SN-LGRB-ABC samples have the smallest intrinsic scatter $\sigma_{Platinum,cor}=\sigma_{SN-LGRB-ABC,cor}=0.22 \pm 0.10$, followed by the KN-SGRB ($\sigma_{KN-SGRB,cor}=0.24 \pm 0.12$) and the gold ($\sigma_{Gold,cor}=0.32 \pm 0.07$) samples. Again, the Type II class is the one with the largest intrinsic scatter even after the correction for selection effects ($\sigma_{Type II,cor} = 0.66 \pm 0.05$). 

%Platinum sample is compatible in 1 $\sigma$ for all parameters. Long and SNe ABC samples are compatible in 1 $\sigma$ for $b$ and $C_0$, while $a$ in 2 $\sigma$. Short is compatible in 2 $\sigma$ for $a$ and $b$, while for $C_0$ is within 3 $\sigma$. XRFs, SNe total and KNe are compatible in 1 $\sigma$ for $a$, 2 $\sigma$ for $b$ and $C_0$. UL is compatible in 1 $\sigma$ for $a$, 2 $\sigma$ for $b$ and 3 $\sigma$ for $C_0$.

We check the compatibility of the gold fundamental plane best-fit parameters ($a$, $b$, and $C_0$ presented in the second half of Table \ref{Table1}) with the other classes after selection biases are taken into account. The platinum, XRF and internal plateau parameters are all compatible in 1 $\sigma$. For the LGRB, SN-LGRB-ABC, and Type II samples, $a$ is compatible in 2 $\sigma$, and $b$ and $C_0$ are compatible in 1 $\sigma$. For the SN-LGRB, ULGRB, and KN-SGRB  samples, $a$ is compatible in 1 $\sigma$, and $b$ and $C_0$ are compatible in 2 $\sigma$. For the sample without internal plateaus and the whole sample, $a$ is compatible in 1 $\sigma$, and $b$ and $C_0$ are compatible in 2 $\sigma$. Lastly, for the SGRB sample there is compatibility in $3$ $\sigma$ for $b$ and $C_0$, and in $1$ $\sigma$ for $a$.

We compute the {\it z}-scores for the evolution (the last two columns of Table \ref{Table2}). Even if the {\it z}-scores change, we reach the same conclusions of the observed samples: the KN-SGRB distribution is still the furthest, with {\it z}-score=10.39, followed by the SN-LGRBs and SGRBs; the conclusions of the SN-LGRBs versus LGRBs and KN-SGRBs versus SGRBs remain unchanged. For the results presented in Table \ref{Table5}, we refer to the Type II as a reference plane, versus Type I, and Type II versus the internal plateau samples. We note that the evolution pushes the {\it z}-scores at around $\mid3\mid$ in both cases. This could be a consequence of the fact that the Gaussian distributions of the distances to the Type II plane have a larger $\sigma$ after the evolutionary effects are considered. The {\it z}-score=0.73 of the ULGRB sample still remains very low, confirming the possibility that ULGRBs and LGRBs may belong to the same physical class. The highest $R^2$ are for SN-LGRB-ABC, SN-LGRB, and KN-SGRB samples=(0.95,0.91,0.90), while the $R^{2}_{adj}$ are for SN-LGRB-ABC, SN-LGRB, and KN-SGRB ones=(0.94, 0.90, 0.87). All the $P$ values remain very low even after correcting for the evolution.

% [FRASE PER SPIEGARE GLI C_O NEGATIVI]. We stress the fact that there is a very strong anticorrelation between parameters $b$ and $C_{0}$ for all the categories. Remembering  that $b$ links L_peak to L_a, this fact can explain why for some classes there is a negative mean value for the normalization $C_{0}$ for L_a.

%In addition, as it is visible from the right panel of Fig. 4, where the probability distributions of smoothed histograms are plotted, the distance of the peaks of the distribution between the gold sample and the SSE is the largest. For details about the definition and how the smoothed histogram has been computed, see Appendix 3. [questa frase va cancellata?]Of note is the fact that the Ultra-long GRBs can be associated with the fundamental plane [lo z score degli UL rimane ancora basso quindi questa conclusione rimane vera]. There are only two Ultra-long GRBs in this sample, so a full analysis of this type will have to wait until a larger sample is available [questa frase non è più vera].

\begin{figure}
\includegraphics[width=0.9\hsize,height=0.9\textwidth,angle=0,clip]{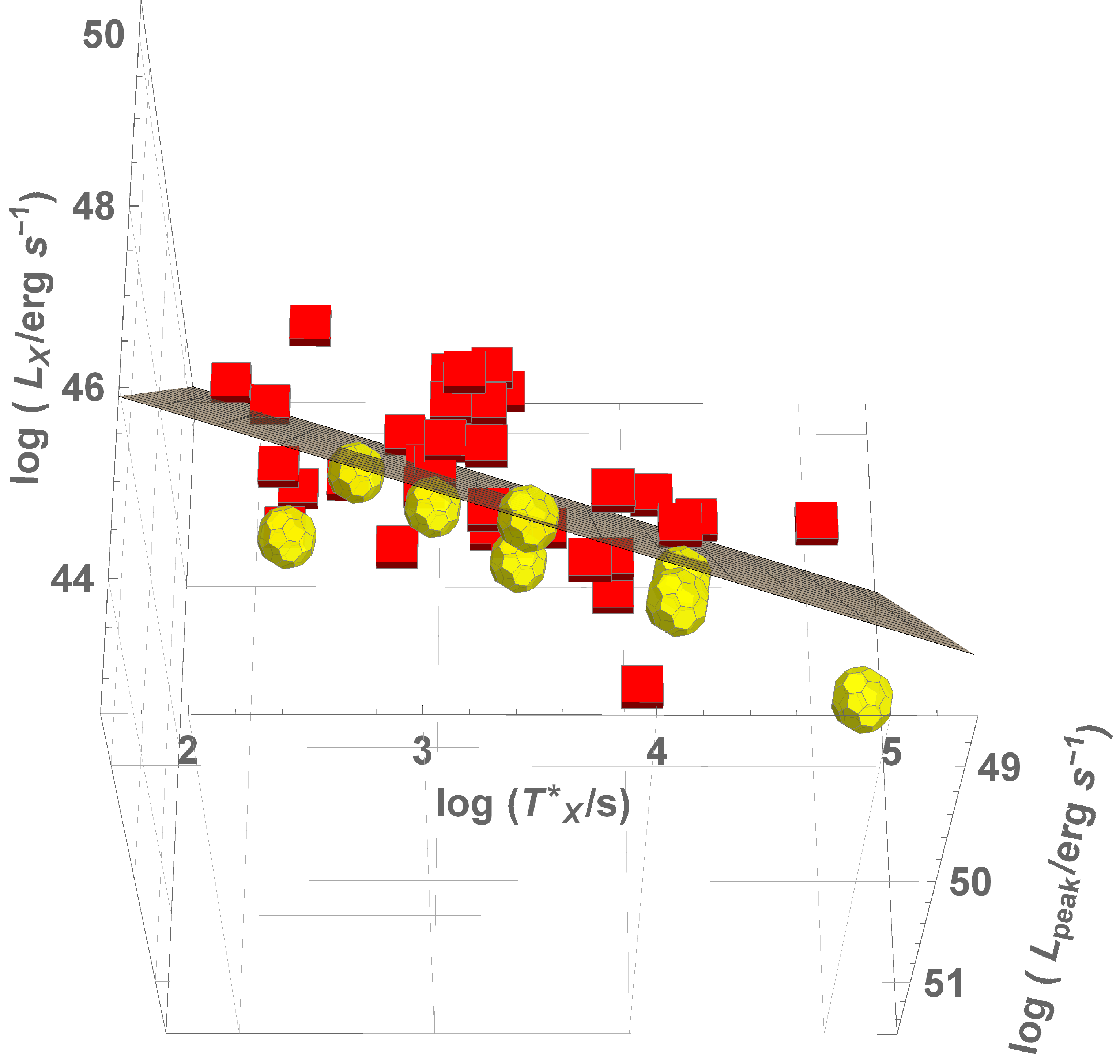}
\caption{The $L_X- T^{*}_{X}-L_{peak}$ relation for the SGRB sample with separated KN-SGRB cases. We note here that all the KN-SGRBs fall below the best-fitting plane.}
\end{figure}
\label{fig6}
\noindent 

\begin{figure}
\includegraphics[width=0.5\hsize,height=0.35\textwidth,angle=0,clip]{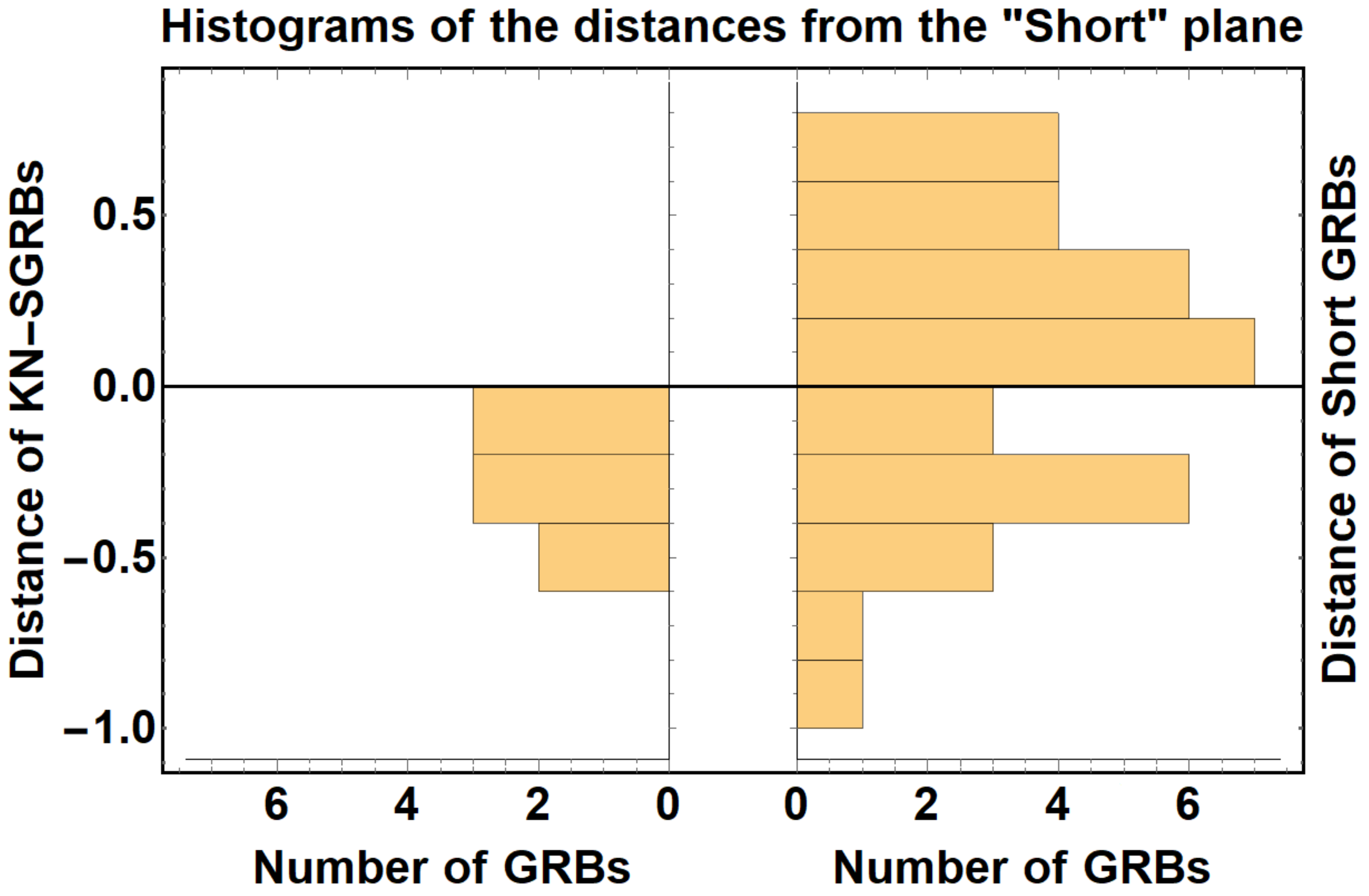}
\includegraphics[width=0.5\hsize,height=0.35\textwidth,angle=0,clip]{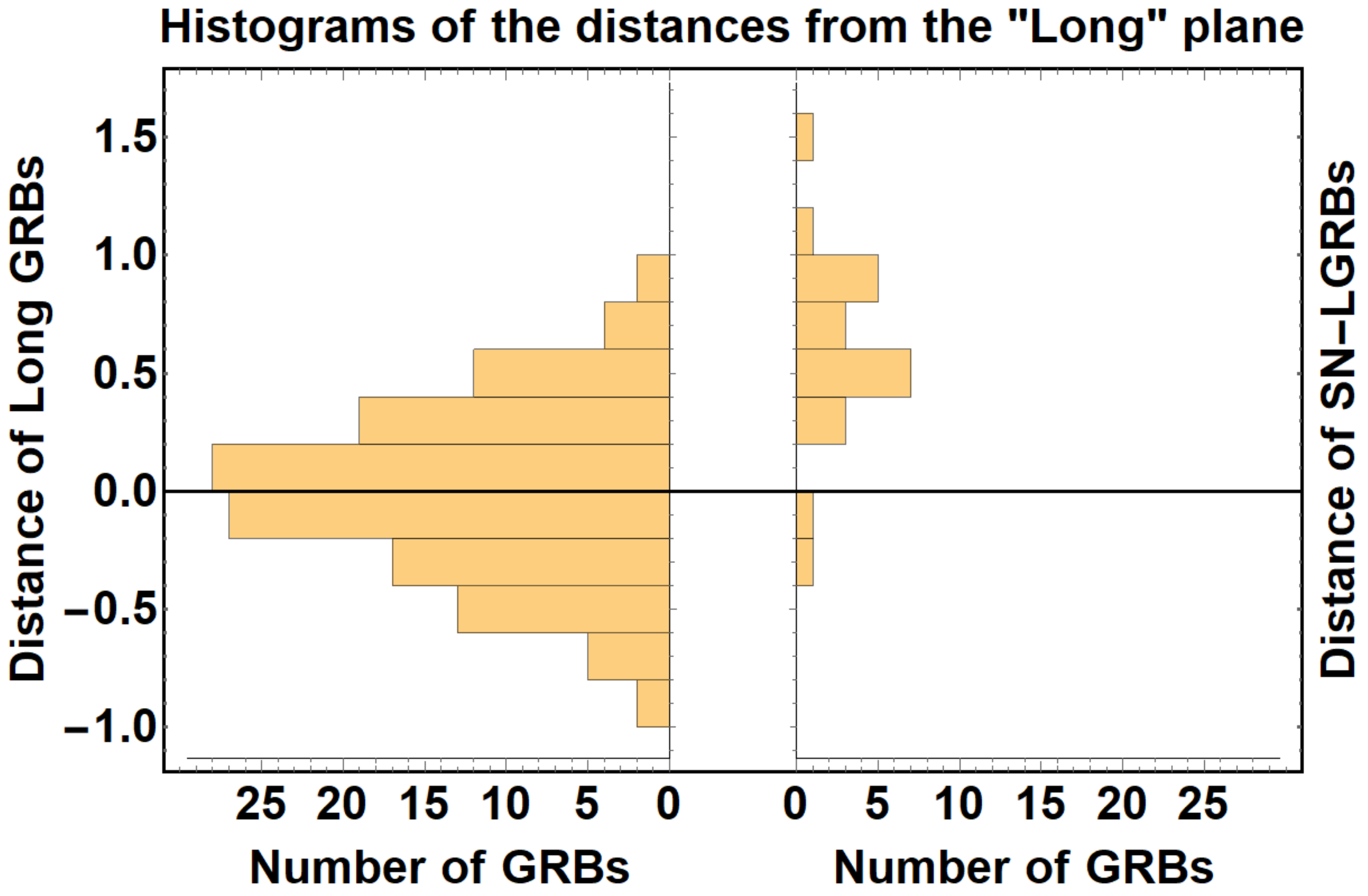}
\includegraphics[width=0.5\hsize,height=0.35\textwidth,angle=0,clip]{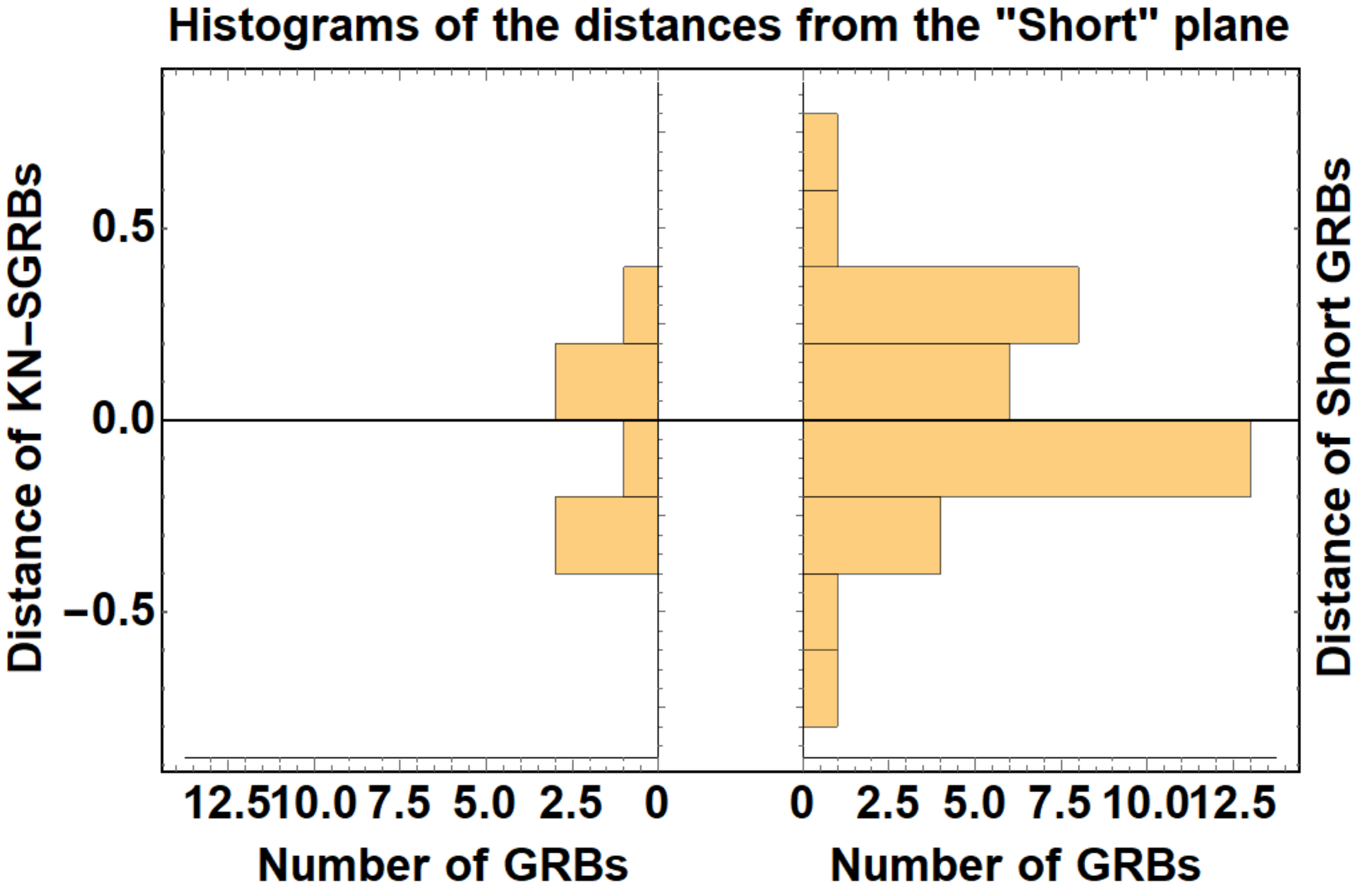}
\includegraphics[width=0.5\hsize,height=0.35\textwidth,angle=0,clip]{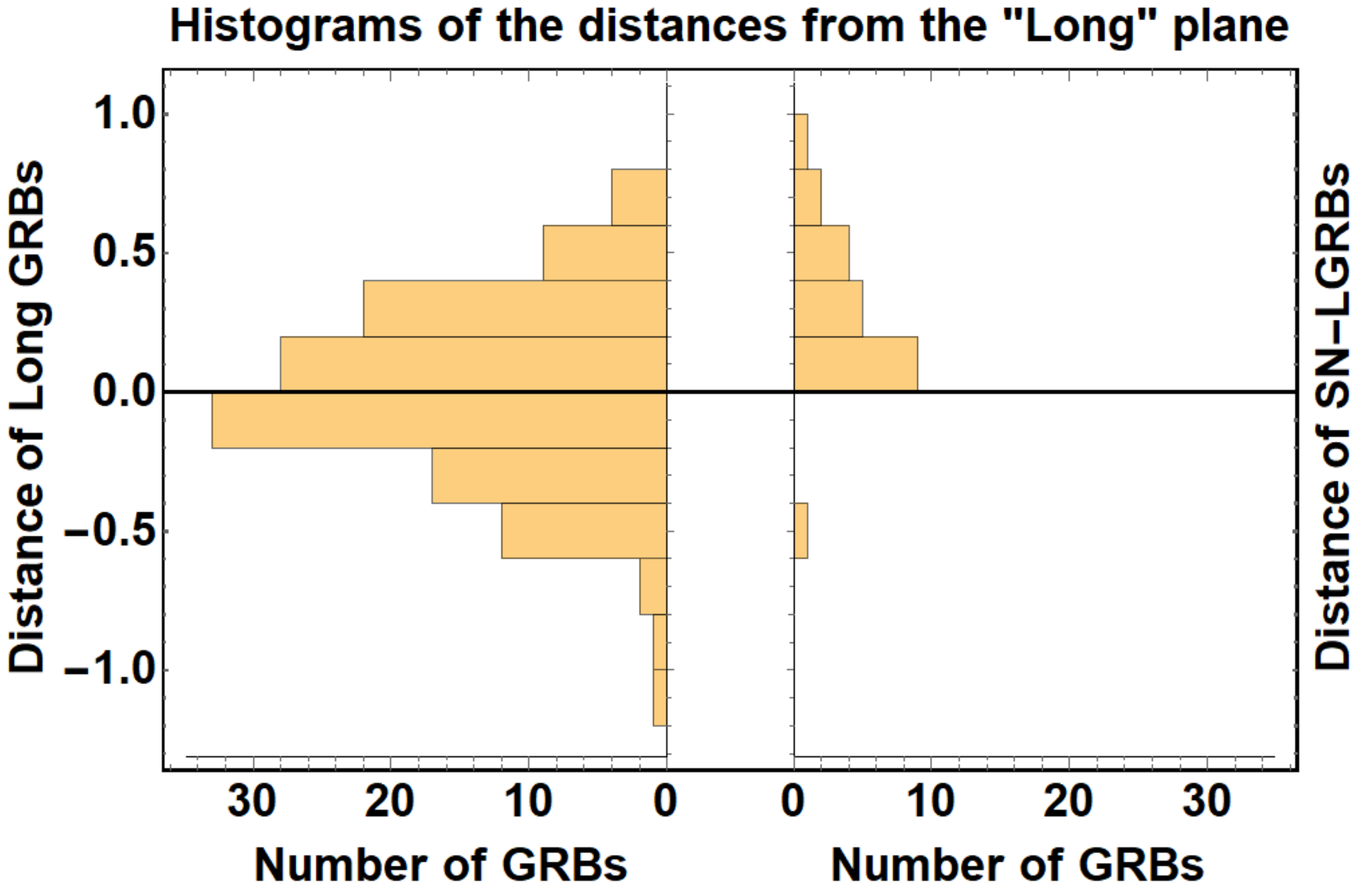}
\caption{Upper left panel shows the histograms of the distance from the short plane for KN-SGRBs and SGRBs, the right upper panel the histograms of the distance from the long plane for LGRBs and SN-LGRBs, both panels take into account the evolution. Lower panel left and right: the same as the upper panels, but without considering the redshift evolution and selection biases.}
\end{figure}
\label{fig7}
\noindent 

\begin{table}
\caption{Table of {\it z}-scores for the SN-LGRB and KN-SGRB Classes with Respect to the LGRB and SGRB Ones Taken as Reference Planes, Respectively, with the Number of GRBs in Each Sample and the Probability that the Fundamental Planes of LGRBs and SN-LGRBs, and KN-SGRBs and SGRBs, are Drawn by the Same Distribution, Respectively}
\begin{tabular}{ |c|c|c|c|c|c|c|c|c| } 
 \hline

Class & $z$-score & $N$ & Probability &  $z$-score cor &
Probability cor\\ \hline
SN-LGRBs & 4.61 & 22 & $\le 10^{-4}$ & 7.08 & $\le 10^{-4}$ \\ 
KN-SGRBs & 0.67 & 8 & 0.50 & 1.91 & 0.06 \\ 
\hline
\end{tabular}

 \textbf{Note.} On the right side of the Table we show {\it z}-scores without evolution, while on the left side the ones with evolution (cor).
\label{Table6}

\end{table}

\section{Discussion and Conclusions}\label{discussion}

In our current investigation of GRB classes, we have enlarged our sample size for all the categories compared to previous works, and we have defined the platinum sample, which reduces the intrinsic scatter given by the updated gold sample of $31.3\%$ once the correction for the selection biases has been taken into account. The KN-SGRBs present small and similar values of $\sigma_{int}$ both with and without considering the evolutionary effects. The stability of the KN-SGRB plane in terms of $\sigma$ and its low value could indicate that GRBs associated with KNe and presenting a plateau can be used as cosmological tools. In particular, it has the third smallest intrinsic scatter, $\sigma_{int}=0.24 \pm 0.12$, after the platinum and the SNe-LGRB-ABC samples, $\sigma_{int}=0.22 \pm 0.10$, when evolutionary effects are considered. Indeed, very recently a study was performed on the use of the kilonovae LCs to constrain the value of $H_0$ (Coughlin et al. 2020).
We have computed the Gaussian fits to the histograms of the distance to the gold fundamental plane from all the classes (see Figure \ref{fig3}) and the {\it z}-score for each category compared to the gold fundamental plane (see Table \ref{Table2}). We have obtained the highest score for the KN-SGRB sample (10.18), followed by the SN-LGRB and SGRB samples, which could indicate different physical mechanisms related to their energy emission. Thus, the fundamental plane relation becomes a crucial tool to discriminate between theoretical models. Interestingly, a very low score has been obtained for the ULGRBs, thus pinpointing the possibility that ULGRBs and LGRBs could come from the same population. The {\it z}-score of the observed SGRB distances from the gold plane here is $33\%$ smaller ($z$-score=5.57) than the one presented in Dainotti et al. (2017a), ($z$-score=8.3), but it still remains significant. We here stress that in this new analysis we have more SGRBs, 43 versus 15 in the previous analysis performed in Dainotti et al. (2017a). In previous analyses the IS were not included, while in our work we have added 12 IS GRBs. 
Most of the parameters obtained, correcting for redshift evolution and selection biases, are consistent within 1 $\sigma$ with the observed ones. %This correction allows a general decrease of $\sigma_{int}$ for all classes \textbf{(the only exception is the ULGRB sample)}: the Gold plane has, \textbf{after this correction}, the smallest $\sigma_{int}$, followed by the Platinum, Long, \textbf{SN-LGRB-ABC} and \textbf{KN-SGRB} planes. 
Thus, the platinum, SNe-LGRB-ABC and KN-SGRB samples are the most suitable candidates to be used as a cosmological standard candle. %\textbf{once the selection effect are taken into account}.

To make the classification more homogeneous from a physical point of view, we have studied the statistical differences of the Gaussian fits to the distances from the Type II fundamental plane from Type I and the internal plateau classes. We find a significant difference between Type I and Type II samples ($z$-scores=$4.28$ without evolution, $3.19$ with evolution), as expected by their possible different nature, while no such difference has been observed between the Type II and internal plateau samples ($z$-scores=$-1.43$ without evolution, $-2.92$ with evolution), as expected by the fact that all the GRBs presenting internal plateau in our samples belong to the Type II class, since in the current sample adopted by us the internal plateau GRBs are all LGRBs, XRFs, and ULGRBs.

We have then studied with the same methods the distances of the KN-SGRBs from the SGRB plane and of the SN-LGRBs from the LGRBs. We find no evident statistical difference between the KN-GRBs and SGRBs ($z$-scores=0.67 without evolution, 1.91 with evolution), but a clear difference for SN-LGRBs and LGRBs ($z$-scores=4.61 without evolution, 7.08 with evolution). For the SGRBs and KN-SGRBs samples, Figure 8 shows that even if there is not a statistical difference in the distance from the short fundamental plane, a clear clustering is present for the KN-SGRBs, that fall all below the short fundamental plane when corrected for evolution and selection biases. All KN-SGRBs are also present below the fundamental plane for the whole sample in both cases with and without considering evolution, while SGRBs are present above and below the fundamental plane for the whole sample in both cases.

Regarding the KNe events, Gompertz et al. (2018) found a difference of 3.5 mag between the KN in SGRB 060614A and the upper limits in SGRB 061201 and 080905A. This may potentially suggest a double binary merger of two NSs (BNS) or an NS and BH merger dichotomy in the SGRB population, as this represents a possible way to explain an apparent contrast in the power ejected by a KN; an NS-BH merger can produce as much as 10 times more dynamical ejecta than a binary NS can (Metzger et al. 2017). Further observations of KNe will reveal whether the magnitude of the emission forms a continuum, or persists to display a gap in brightness between the two populations. If the latter hypothesis is revealed to be true, then we will divide the fainter and brighter events and analyze their planes separately.

The increase of the KNe sample will occur with future observations from Swift and from future satellites such as SVOM (Mate et al. 2019 and Wei et al. 2016, which will be launched in 2021) and THESEUS (Amati et al. 2018, Stratta et al. 2018), which if approved will be launched in 10 yr. The possibility to further confirm the reliability of the KN-SGRB and SN-LGRB fundamental planes relation as a tool both to discriminate between theoretical models and for cosmological applications is encouraging us to pursue further studies in both directions.

\section{acknowledgement}
This work made use of data supplied by the UK Swift Science Data Centre at the University of Leicester. We are grateful to S. Savastano for helping write the python codes. We are grateful to G. Srinivasaragavan, R. Wagner, L. Bowden, Z. Nguyen, and R. Waynne for help with the lightcurve parameter fitting. M.G.D. ackowledges support from the American Astronomical Society Chretienne Fellowship and from Miniatura 2 and the Department of Energy of State who funded the summer internship of Srinivasaragavan, R. Wagner, L. Bowden, and R. Waynne.
S.N. is partially supported by JSPS Grants-in-Aid for Scientific Research KAKENHI (A) 19H00693", Pioneering Program of RIKEN for Evolution of Matter in the Universe (r-EMU), and Interdisciplinary Theoretical and Mathematical Sciences Program 
(iTHEMS) of RIKEN. N. F. acknowledges financial support from UNAM-DGAPA-PAPIIT through grant IA102019. 
%\bibliography{sample63}{}

\begin{thebibliography}{99}

\bibitem[Abbott et al. (2017)]{Abbott} Abbott, B. P., Abbott, R., Abbott, T., D., et al., 2017, Phys. Rev. L., 119, 16, 161101;

\bibitem[Amati (2018)]{Amati} Amati, L., O'Brien, P., G\"otz, D., et al., 2018, ASR, 62, 1;

\bibitem[Avni (1976)]{Avni1976}
Avni, Y. 1976, ApJ, 210, 642;

%\bibitem[Barthelmy et al. 2005] {Barthelmy}  Barthelmy, S. D., Barbier, L., M.,  et al., 2005, SSRv., 120, 3, 143; 

\bibitem[Berger et al. (2013)]{Berger2013} Berger, E., Fong W.,  Chornock R., 2013, ApJL, 774, 2, L23;

\bibitem[Berger (2014)]{Berger}  Berger, E.,  2014, A \& A, 52, 43-105;

\bibitem[Bloom et al. (1999)]{Bloom1999} Bloom, J. S., Kulkarni S. R., Djorgovski, S. G., et al., 1999, Nature, 401, 6752;

\bibitem[Bloom et al. (2001)]{Bloom} Bloom, J. S., Frail, D., A., Sari, R., 2001, ApJ, 121, 6, pp. 2879-2888;

\bibitem[Chevalier et al. (2000)]{Chevalier}  Chevalier, R. A. \& Li, Z., 2000, ApJ, 536, 1, 195-212;

\bibitem[Coughlin et al. (2020)]{Coughlin} Coughlin, M. W., Antier, S., Dietrich, T., et al., 2020, Nature Communications, 11, id. 4129;


%\bibitem[Burrows et al. 2006] {Burrows} Burrows, D. N., Hill J. E., et al., 2005, SSRv, 120, 3, 165;
\bibitem[Coulter et al. (2017)]{Coulter} Coulter, D. A., Foley, R., J., Kilpatrick, C. D., et al., 2017, Science, 358, 6370;

\bibitem[D'Agostini (2005)]{Dago05} D'Agostini, G., 2005, arXiv:physics/0511182;


\bibitem[Dai et al. (1998)]{Dai}  Dai, Z. G. \& Lu, T., 1998, Phys Rev. Letters, 81, 20, 4301-4304;


\bibitem[Dainotti et al. (2008)]{Dainotti2008}  Dainotti, M. G., Cardone, V. F., Capozziello, S., 2008, MNRAS, 391, 1,  L79–L83;

\bibitem[Dainotti et al. (2010)]{Dainotti2010}
Dainotti, M.G., Willingale, R., Capozziello S., Fabrizio Cardone, V., Ostrowski, M., ApJL, 722, L215.

\bibitem[Dainotti et al. (2013)]{Dainotti2013}
Dainotti, M. G., Petrosian, V., Singal, J., Ostrowski, M., 2013, ApJ, 774, 157;

\bibitem[Dainotti et al. (2015a)]{Dainotti2015a}
Dainotti, M. G., Del Vecchio, R., Nagataki, S., Capozziello, S., 2015, ApJ, 800, 31;

\bibitem[Dainotti et al. (2015b)]{Dainotti2015b}
Dainotti, M. G., Petrosian, V., Willingale, R., et al., 2015, MNRAS, 451, 4;

\bibitem[Dainotti et al. (2016)]{Dainotti2016}
Dainotti, M. G., Postnikov, S., Hernandez, X., Ostrowski, M., 2016, ApJL, 825, 2, id L20, 6

\bibitem[Dainotti et al. (2017a)]{Dainotti2017a}  Dainotti, M. G., Hernandez, X., Postnikov, S., et al., 2017a, ApJ, 848, 88, 2017;

\bibitem[Dainotti et al. (2017b)]{Dainotti2017b}
Dainotti, M. G., Nagataki, S., Maeda, K., Postnikov, S., \& Pian, E., 2017b, A\&A, 600A, 98.

\bibitem[Del Vecchio et al. (2016)]{Del Vecchio}
Del Vecchio, R., Dainotti, M.G., Ostrowski, M., 2016, ApJ, 828, 1, id 36, 6

\bibitem[Eddington (1913)]{Eddington1913}
Eddington, A. S., 1913, MNRAS, 73, 359-360;

\bibitem[Eddington (1940)]{Eddington1940}
Eddington, A. S., Sir, 1940, MNRAS, 100, 354;

\bibitem[Efron \& Petrosian (1992)]{Efron}
Efron, B., Petrosian, V., 1992, ApJ, 399,  p.345;

%\bibitem[Eichler et al. 1989] {Eichler}  Eichler, D., Livio, M., Piran, T., Schramm, D. N., 1989, Nature, 340, 126-128;

\bibitem[Evans et al. (2009)]{Evans}
Evans, P. A., Beardmore, A.P., Page, K.L., et al. 2009, MNRAS, 397, 3, 1177;

\bibitem[Fraija et al. (2020)]{Fraija} Fraija, N.,  Betancourt Kamenetskaia, B.,  Dainotti, M. G, et al., 2020, arXiv:2006.04049;

%\bibitem[Galama et al. 1998 ]{Galama} Galama T. J., Vreeswijk, P. M., et al., 1998, Nature, 395, 6703;

\bibitem[Gao et al. (2015)]{Gao2015}  Gao, H., Ding, X., Wu, X., Dai, Z., Zhang, B., 2015, ApJ, 807, 2, 163;

\bibitem[Gao et al. (2017)]{Gao2017} Gao, H., Zhang, B., L\"{u}, H., Li, Y., 2017, ApJ, 837, 1, 50;

\bibitem[Gendre et al. (2013)]{Gendre2013}  Gendre, B., Stratta, G., Atteia, J. L., et al., 2013, ApJ, 766, 30, 1;

\bibitem[Gendre et al. (2019)]{Gendre2019}  Gendre, B., Joyce Q. T., Orange, N. B., et al., 2019, MNRAS, 486, 2;

\bibitem[Goldstein et al. (2017)]{Goldstein} Goldstein, A., Veres, P., Burns, E., et al., 2017, ApJL, 848, L14;

\bibitem[Gompertz et al. (2018)]{Gompertz2018} Gompertz B. P., Levan A. J., Tanvir, N. R., et. al., 2018, ApJ, 860, 1, id 62;

\bibitem[Hjorth \& Bloom (2011)]{Hjorth2011}
Hjorth, J., \& Bloom, J.S, 2011, `Gamma-Ray Bursts", eds. C. Kouveliotou, R. A. M. J. Wijers, S. E. Woosley, Cambridge University Press, 2011.

%\bibitem[Jin et al. 2016] {Jin2016} Jin, Z., Hotokezaka, K., et al., 2016, Nature communications, 7;

%\bibitem[Jin et al. 2018] {Jin2018} Jin, Z., Li, X., et al., 2018, ApJ, 857, 2;

%\bibitem[Kasliwal et al. 2017] {Kasliwal} Kasliwal, M. M., Nakar, E., et al., 2017, ApJL, 843, L34;

\bibitem[Kann et al. (2011)]{Kann} Kann, D. A., Klose, S., Zhang, B., et al., 2011, ApJ, 734, 2, 96;

\bibitem[Kouveliotou et al. (1993)]{Kouveliotou}
Kouveliotou, C., Meegan, C. A., Fishman, G. J., et al., 1993, ApJ, 413, 2, L101;

\bibitem[Lattimer \& Schramm (1976)]{Lattimer}  Lattimer, J. M. \& Schramm, D. N., 1976, ApJ, 210, 1, 54;

\bibitem[Levan et al. (2007)]{Levan07} 
Levan, A. J., Jakobsson, P., Hurkett, C., et al., 2007, MNRAS, 378, 4, 1439

\bibitem[Levesque et al. (2010)]{Levesque} Levesque, E. M., Bloom J. S., Butler, Nathaniel R., et al., 2010, MNRAS, 401, 963;

%\bibitem[Li et al 2012.] {Li}  Li, L., Liang E. W., et al., 2012, ApJ, 758, 27;

%\bibitem[MacFadyen et al., 2001] {MacFadyen}  MacFadyen, A., Woosley, S. E., Heger A., 2001, ApJ, 550, 1, 410;
\bibitem[Li et al. (2018)]{Li2018}  Li, L., Wu, X., Lei, W., et al., 2018, ApJS, 236, 2, 26;


\bibitem[Li et al. (2020)]{Li} Li, Y., Zhang, B., Yuan, Q., 2020, ApJ, 897, 2, 154;

\bibitem[Lyons et al. (2010)]{Lyons} Lyons, N., O'Brien, P. T.; Zhang, B., et al., 2010, MNRAS, 402, 2, 705-712;

\bibitem[Malmquist (1925)]{Malmquist} Malmquist, K. G., 1925, Meddelanden fran Lunds Astronomiska Observatorium Series I, 106, 1-12;

\bibitem[Mate et al. (2019)]{Mate} Mate, S., Bouchet, L., Atteia, J. L., et al., 2019, Experimental Astronomy, 48, 2-3;

\bibitem[Mazets et al. (1981)]{Mazets}
Mazets, E. P., Golenetskii, S. V., Ilyinskii,V. N., et al., 1981, Astrophys. Space Sci., 80, 1, 3;

\bibitem[Metzger \& Berger (2012)]{Metzger} Metzger, B. D. \& Berger E., 2012, ApJ, 746, 48;

\bibitem[Metzger et al. (2017)]{Metzger2017} Metzger, B. D.,  Living Reviews in Relativity, 20, 3;

\bibitem[Narayan et al. (1992)]{Narayan}  Narayan, R., Paczynski, B., Piran, T., 1992, ApJL, 395, 2, 83;

\bibitem[Norris et al. (2006)]{Norris}  Norris, J. P. \& Bonnell, J. T., 2006, ApJ, 643, 1, 266;

\bibitem[Norris et al. (2010)]{Norris2010}
Norris, J.P., Gehrels, N. \& Scargle, J. 2010, ApJ, 717, 411.

\bibitem[O'Brien et al. (2006)]{O'Brien}  O'Brien, P. T., Willingale, R., Osborne, J., et al., 2006, ApJ, 647, 2, 1213;

%\bibitem[Panaitescu \& Vestrand 2008] {Panaitescu2008}  Panaitescu, A. \& Vestrand, W. T.,2008, MNRAS, 387, 2, 497-504;

%\bibitem[Panaitescu \& Vestrand 2011]{Panaitescu2011}  Panaitescu, A. \& Vestrand, W. T., 2011, MNRAS, 414, 4, 3537;

%\bibitem[Pescalli et. al. (2015)]{Pescalli2015}
%Pescalli, A., Ghirlanda, G., Salafia, O.S., et al. 2015, MNRAS, 447, 1911

\bibitem[Petrosian et al. (2015)]{Petrosian} Petrosian, V., Kitanidis, E., Kocevski, D., 2015, ApJ, 806, 1, 44;

\bibitem[Piro et al. (2014)]{Piro} Piro, L., Troja, E., Gendre, B., et al. 2014, ApJ, 790, L15, 5;

%\bibitem[Roming et al. 2005] {Roming}  Roming P. W. A., Kennedy, T. E., et al., 2005, SSRv, 120, 3-4, 95; 

\bibitem[Rossi et al. (2020)]{Rossi} Rossi, A., Stratta, G., Maiorano, E., et al., 2020, MNRAS, 493, 3;


\bibitem[Sakamoto (2007)]{Sakamoto2007} Sakamoto, T., Hill J. E., Yamazaki, R, et al., 2007, ApJ, 669, 2, 1115;

\bibitem[Sakamoto (2010)]{Sakamoto2011}
Sakamoto, T., Barthelmy, S.D., Baumgartner, W.H., et al. 2011, ApJS, 195, 2.


%\bibitem[Schaefer (2007)]{Schaefer2007}
%Schaefer, B. 2007, ApJ, 660, 16.

\bibitem[Scolnic et al. (2018)]{Scolnic} Scolnic, D. M., Jones, D. O., Rest, A., et al., 2018, ApJ, 859, 101;
%\bibitem[Si et al. 2018]{Si}  Si, S., Qi, Y.Q., et al., 2018, ApJ 863, 50;

\bibitem[Srinivasaragavan et al. (2020)]{Gokul2020} Srinivasaragavan, G., Dainotti, M. G., Fraija, N., et al., 2020, ApJS accepted.


%\bibitem[Stratta et al. 2013] {Stratta2013}  Stratta, G., Gendre, B., et al., 2013, ApJ, 779, 66, 1;

\bibitem[Stratta et al. (2018)]{Stratta2018} Stratta, G., Ciolfi, R., Amati, L., et al., 2018, ASR, 62,  3;

%\bibitem[Tanvir et al. 2013] {Tanvir2013} Tanvir N. R., Levan, A. J. et al., 2013, Nature, 500, 7464, 547-549;


%\bibitem[Tanvir et al. 2017]{Tanvir2017} Tanvir N. R., Levan A. J., et al., 2017, ApJL, 848, L27;

%\bibitem[Troja et al. 2016b]{Troja2016b} Troja, E., Sakamoto, T., et al., 2016, ApJ, 827, 102;
\bibitem[Wei et al. (2016)]{Wei} Wei, J., Cordier, B., Antier S., et al., 2016, arXiv:1610.06892.


\bibitem[Willingale et al. (2007)]{W07}
Willingale, R., O'Brien, P. T., Osborne, J. P., et al., 2007, ApJ,  662, 1093.

\bibitem[Woosley et al. (1993)]{Woosley} Woosley, S. E., Langer, N., Weaver, T. A., 1993, ApJ, 405, 1, 273;


\bibitem[Virgili et al. (2013)]{Virgili} Virgili, F. J.,  Mundell, C. G., Pal'shin, V., et al., 2013, ApJ, 778, 54;

\bibitem[Xiao \& Schaefer (2009)]{Xiao} 
Xiao, L. \& Schaefer, B.E. 2009, ApJ, 707, 387.

\bibitem[Yang et al. (2015)]{Yang} Yang, B., Jin, Z., Li, X., et al., 2015, Nature communications, 6,  id. 7323;

\bibitem[Zhang et al. (2006)]{Zhang2006} Zhang, B., Fan, Y. Z.,  Dyks, J., et al., 2006, ApJ, 642, 1, 354;

\bibitem[Zhang et al. (2009)]{Zhang2009}  Zhang, B., Zhang, B., Virgili, F. J.,  et al., 2009, ApJ, 703, 2;


\bibitem[Zhang et al. (2014)]{Zhang2014} Zhang, B. B., Zhang, B., Murase, K.,  Connaughton, V., Briggs, M. S.,  2014, ApJ, 787, 66;

\end{thebibliography}

\bibliographystyle{aasjournal}



\end{document}